 \documentclass[onecolumn,11pt]{IEEEtran}

\usepackage{amsmath, amsfonts, amssymb}
\usepackage{graphicx}
\usepackage{multirow}

\def\pr#1{\mathbb{P}\left(#1\right)}

\begin{document}

\title{Capacity and Delay Analysis of Next-Generation
   Passive Optical Networks (NG-PONs) -- Extended Version\thanks{Technical 
Report, School of Electrical, Computer, and Energy Eng., Arizona
State University, November 2010. 
This extended technial report accompanies~\cite{AuSRGM11}.}}

\author{Frank Aurzada, Michael Scheutzow,
\thanks{F. Aurzada and M.\ Scheutzwo are with the Department of Mathematics,
Technical University Berlin,
10623 Berlin, Germany,
Email: \{aurzada, ms\}@math.tu-berlin.de}
Martin Reisslein,
\thanks{M.~Reisslein is with the Department of Electrical Engineering,
Arizona State University, Tempe,
Arizona 85287-5706, USA,
Email: reisslein@asu.edu}
Navid Ghazisaidi, and Martin Maier
\thanks{N.~Ghazisaidi and M.~Maier are with the Optical Zeitgeist Laboratory,
INRS, University of Qu\'ebec,
Montr\'eal, QC, H5A 1K6, Canada,
Email: \{navid,maier\}@emt.inrs.ca}}

\maketitle

\begin{abstract}
Building on the Ethernet Passive Optical Network (EPON) and Gigabit
PON (GPON) standards, Next-Generation (NG) PONs $(i)$ provide
increased data rates, split ratios, wavelengths counts, and fiber
lengths, as well as $(ii)$ allow for all-optical integration of
access and metro networks. In this paper we provide a comprehensive
probabilistic analysis of the capacity (maximum mean packet
throughput) and packet delay of subnetworks that can be used
to form NG-PONs. Our analysis can cover a wide range of NG-PONs
through taking the minimum capacity of the subnetworks making up the
NG-PON and weighing the packet delays of the subnetworks. Our
numerical and simulation results indicate that our analysis quite
accurately characterizes the throughput-delay performance of
EPON/GPON tree networks, including networks upgraded with higher
data rates and wavelength counts. Our analysis also characterizes
the trade-offs and bottlenecks when integrating EPON/GPON tree
networks across a metro area with a ring, a Passive Star Coupler
(PSC), or an Arrayed Waveguide Grating (AWG) for uniform and
non-uniform traffic. To the best of our knowledge, the presented
analysis is the first to consider multiple PONs interconnected via a
metro network.
\end{abstract}




\baselineskip  0.26in

\section{Introduction}
\vspace{-0.2cm}
The Passive Optical Network (PON) is one of the most widely deployed
access networks due to its unique benefits,
including transparency against data rate and signal format.
The two major state-of-the-art PON
standards IEEE 802.3ah Ethernet PON (EPON) and ITU-T G.984
Gigabit PON (GPON) consist both of a single upstream wavelength channel
and a separate single downstream wavelength channel, whereby both channels are
operated with time division multiplexing (TDM).
EPON and GPON are expected to coexist
for the foreseeable future as they evolve into
Next-Generation PONs (NG-PONs)~\cite{Effe07,KaSGW07,ZhALEY09}.
NG-PONs are mainly envisioned to
$(i)$ achieve higher performance parameters, e.g.,
higher bandwidth per subscriber, increased split ratio,
and extended maximum reach, than current EPON/GPON architectures~\cite{Lin08},
and $(ii)$ broaden EPON/GPON functionalities to include, for instance,
the consolidation of optical access, metro, and backhaul networks, the support of topologies other
than conventional tree structures, and protection~\cite{GrEl08}.
Throughout, network operators are seeking
NG-PON solutions that can transparently coexist with legacy PONs on the
existing fiber infrastructure and enable gradual upgrades
in order to avoid
costly and time consuming network modifications and stay flexible for
further evolution paths~\cite{Effe07}.

In this paper, we evaluate the
capacity (maximum mean packet throughput)
and packet delay of a wide range of NG-PONs through probabilistic analysis
and verifying simulations.
More specifically, we analyze the capacity and delay of various subnetworks
from which NG-PONs can be formed, thus enabling analytical capacity and delay
characterization for a wide range of NG-PONs built from the examined
subnetworks.
Two important applications for our analysis are: (A) The obtained results
provide insight into the performance limitations of
candidate NG-PON architectures and thus inform network operators seeking to
upgrade their installed TDM PONs.
(B) Neither IEEE 802.3ah EPON nor ITU-T G.984 GPON
standardizes a specific dynamic bandwidth allocation (DBA)
algorithm. The design of DBA algorithms is left to manufacturers which
aim at equipping network operators with programmable DBA algorithms
that adapt to
new applications and business models and thus make PONs future-proof.
Our capacity and delay analysis provides an upper throughput bound
and a delay benchmark for gated service~\cite{KMP0202}
which can be used to evaluate the
throughput-delay performance of current and future DBA
algorithms for NG-PONs.

This paper is structured as follows. In the following section, we
review related work on the analysis of PON access and metro packet
networks. In Section~\ref{sec:EPON/GPON} we give overviews of the
EPON and GPON access networks. In Section~\ref{sec:NG-PONs}, we
present NG-PONs that either $(i)$ upgrade PONs or $(ii)$
interconnect multiple PONs across a metropolitan area. We conduct
the capacity and delay analysis of the subnetworks making up NG-PONs
in Sections~\ref{analysis:sec} and~\ref{sec:analysisdel}. 
We first introduce the network model
and then evaluate the capacity (maximum mean aggregate throughput).
Subsequently, we analyze the packet delays. In
Section~\ref{num:sec}, we compare numerical throughput-delay results
obtained from our analysis with simulations and illustrate the
application of our capacity analysis to identify bottlenecks in
NG-PONs. We briefly summarize our contributions in
Section~\ref{sec:conclusions}.

\section{Related Work}
\label{lit:sec}
In this section we briefly review related work on
the analysis of passive optical networks and metropolitan
area networks.
EPONs employ medium access control with an underlying polling
structure~\cite{AYDA1103,MZC0303,NM0806,ZhMo09}.
Building directly on the extensive
literature on polling systems, see e.g.,~\cite{Tak86},
Park et al.~\cite{PaHR05}
derive a closed form delay expression for a
single-channel EPON model with random independent
switchover times.
The EPON model with independent switchover times
holds only when successive upstream transmissions are separated by a
random time interval sufficiently large to ``de-correlate''
successive transmissions, which would significantly reduce bandwidth
utilization in practice.

In an EPON,
the service (upstream transmission)
of an Optical Network Unit (ONU) follows immediately (separated by a guard
time) after the upstream transmission of the preceding ONU
to ensure high utilization.
The switchover time is therefore generally highly
dependent on the round-trip delays and the
masking of the round-trip delays through the interleaving
of upstream transmissions~\cite{KMP0202}.
Subsequent analyses have strived to model these correlated
transmissions and switchovers with increasing fidelity.
In particular, EPONs with a static bandwidth allocation
to the ONUs were analyzed in~\cite{Holmb06,LaVC07} and it was found that
the static bandwidth allocation can meet
delay constraints only at the expense of low network utilization.
Packet delay analyses
for single-channel EPONs with dynamic bandwidth allocation
have been undertaken
in~\cite{AuSHMR08,BaSY07,BaS09,BhGB06,LaVC07,NgGB08,VaLo09}.
A dynamic bandwidth allocation scheme with traffic prediction
assuming a Gaussian prediction error distribution
was analyzed in~\cite{LuAn05icc}.
A grant estimation scheme
was proposed and its delay savings analyzed
in~\cite{ZhMa08}.

GPONs have received relatively less research interest than EPONs.
To the best of our knowledge we conduct the first delay analysis
of GPONs in this paper.

Similarly, WDM PONs have received relatively little research
attention to date.
The call-level performance of a WDM PON employing
Optical Code Division Multiple Access was analyzed in~\cite{VaVL09}.
To the best of our knowledge a
packet-level analyses of WDM EPONs has so far only been
attempted in~\cite[Section~2.4]{Chang08} where an
offline scheduling WDM EPON was analyzed with the help of a two stage queue.
In offline EPONs, also referred to as EPONs with interleaved polling
with stop~\cite{ZhMo09},
the Optical Line Terminal (OLT) collects bandwidth requests from
all ONUs before making bandwidth allocation decisions.
An offline scheduling EPON was also considered in~\cite{AuSRM10}
where the stability limit (capacity) and packet delay were analyzed.
In contrast, we analyze in this paper the capacity and packet delay
for WDM PON channels that are integrated into an online scheduling
PON, i.e., transmissions from the different ONUs are interleaved
to mask propagation delays.

Metropolitan area networks have received significant attention over
the past two decades. The capacity and delay performance of packet
ring networks with a variety of MAC protocols has been analyzed in
numerous studies, see e.g.,~\cite{CaFC00,MBLM+96,MLMN00,RW96,ScRMS08}. More
recently, analyses of medium access and fairness mechanisms for the
resilient packet ring have been undertaken, see for
instance~\cite{DaGj04,GYBL04}. The impact of fiber shortcuts in ring
networks has been analyzed in~\cite{RH97}. Metropolitan star
networks based on passive star couplers (see e.g.,~\cite{Mehr90})
and arrayed waveguide gratings 
(see e.g.,~\cite{CHNS00,MaRe04,ScMRW03,YaMRC03}) 
have been
analyzed in isolation. In this paper, we analyze to the best of our
knowledge for the first time a comprehensive NG-PON that
interconnects multiple PONs via a metro network combining a ring and
star networks.

\section{Overview of EPON and GPON}
\label{sec:EPON/GPON}
While EPONs are deployed mostly in the Asia-Pacific region, GPONs are leading in the U.S. and Europe. Typically, both EPON
and GPON have a physical tree topology with the Optical Line Terminal
(OLT) at the root. The OLT connects through an optical
splitter to multiple Optical Network Units (ONUs), also known as
Optical Network Terminals (ONTs).
Each ONU can serve a single or multiple
subscribers. To facilitate DBA, both EPON and GPON use polling based
on a \emph{report/grant mechanism}. In each polling cycle, ONUs
send their instantaneous upstream bandwidth demands through report
messages to the OLT, which in turn dynamically allocates variable
upstream transmission windows by sending a separate grant message to
each ONU. EPON and GPON have some major and minor differences with
respect to polling timing structure,
line rate, reach, split ratio, guard time, protocol
overhead, and bandwidth efficiency~\cite{Effe07}.

\subsection{Major Differences}
The EPON is a symmetric network providing a data rate of 1 Gb/s in both
upstream and downstream directions. It provides a reach between OLT
and ONUs of up to 20 km for a split ratio as high as 1:64.
The EPON has variable-length polling cycles based on the
bandwidth demands, which are signalled with the
multipoint control protocol (MPCP). The ONU uses the MPCP REPORT
message to report bandwidth requirements of
up to eight priority queues to the OLT. The OLT passes received REPORT
messages to its DBA algorithm module to calculate the upstream
transmission schedule.
Then, the OLT issues upstream transmission grants by
transmitting a GATE message to each ONU.
Each GATE message supports up to four transmission grants,
each specifying the start time and length of the transmission
window. The transmission window may comprise multiple Ethernet frames,
whereby EPON does not allow for fragmentation. EPON carries Ethernet
frames natively, i.e., without encapsulation.

The GPON offers several combinations of upstream/downstream data rates
with a maximum symmetric data rate of 2.488 Gb/s. GPON supports up to
60 km reach, for a maximum split ratio of 1:128.
Both upstream and downstream transmissions are
based on a periodically recurring time structure with a fixed frame
length of $\delta = 125\ \mu$s.
Each upstream frame contains
dynamic bandwidth report (DBRu) fields.
Each downstream frame contains a physical
control block (PCBd), which includes
a bandwidth map (BWmap) field specifying the ONU upstream
transmission grants.
Unlike the EPON, the GPON
deploys the GPON encapsulation method (GEM) involving a 5-byte GEM
header and allows for Ethernet frame fragmentation. Two DBA methods
are defined for GPON: ($i$) status-reporting DBA based on
ONU reports via the DBRu field, and ($ii$) non-status-reporting DBA
based on traffic monitoring at the OLT. Recent GPON research
has focused on the design and evaluation
of status-reporting DBA algorithms~\cite{ChKS06,JiHS06,JiSe09}.

\subsection{Minor Differences}
\label{overheads:sec}
EPON uses various guard times between two neighboring transmission
windows, e.g., the laser on-off time. In GPON,
additional fields are used in each upstream and downstream frame,
e.g., the physical layer operation, administration, and maintenance
(PLOAM) field. The impact of the various types of guard times and
overhead fields on the bandwidth efficiency of both EPON and GPON was
thoroughly investigated in~\cite{HaSM06,SCAW09}, and we consequently
neglect all overheads in our analysis to uncover the
fundamental underlying performance characteristics due to the different
polling timing structures, i.e., variable-length cycles in EPON and
fixed-length frames in GPON.

\section{NG-PONs}
\label{sec:NG-PONs}
NG-PONs are PONs that provide $(i)$ higher data rates, larger counts of
wavelength channels, longer ranges, and/or higher split ratios,
as well as $(ii)$ broader functionalities than
current EPON and GPON networks, as explained next.

\subsection{High-speed TDM PON}
Higher speeds are needed to
support emerging bandwidth-hungry applications, e.g.,
high-definition television and video on demand, and to provide sufficient
capacity as backhauls of next-generation IEEE 802.11n wireless LANs
with a throughput of 100 Mb/s or higher per device~\cite{Lin08}.
For both EPON and GPON, standardization efforts
have begun to specify symmetric or asymmetric data rates of up to 10
Gb/s~\cite{Effe07}. DBA algorithms for EPON, GPON,
and high-speed TDM PON are compared in~\cite{SCAW09}.

\subsection{WDM PON}
\label{sec:WDM PON}
Different forms of WDM PONs have been
actively studied as a component of NG-PON~\cite{GrEl08}.
In a wavelength-routing WDM PON, each ONU is assigned a
dedicated pair of wavelength channels for upstream and downstream
transmission, which brings some advantages, but requires
replacing the power splitter in installed TDM
PONs with a wavelength demultiplexer.
According to~\cite{Effe07}, a more practical
approach towards WDM PONs is to leave the existing power-splitting PON
infrastructure in place and to select wavelengths at each
ONU using a bandpass filter (BPF) with a small insertion loss of 1
dB. To ensure that WDM enhanced ONUs, operating on additional
wavelengths, can be installed on legacy TDM PON infrastructures, the
conventional TDM ONUs may be equipped
with wavelength blocking filters which let only the legacy TDM
wavelength pass.
\begin{figure*}[t]
\begin{center}
\includegraphics[width=\textwidth]{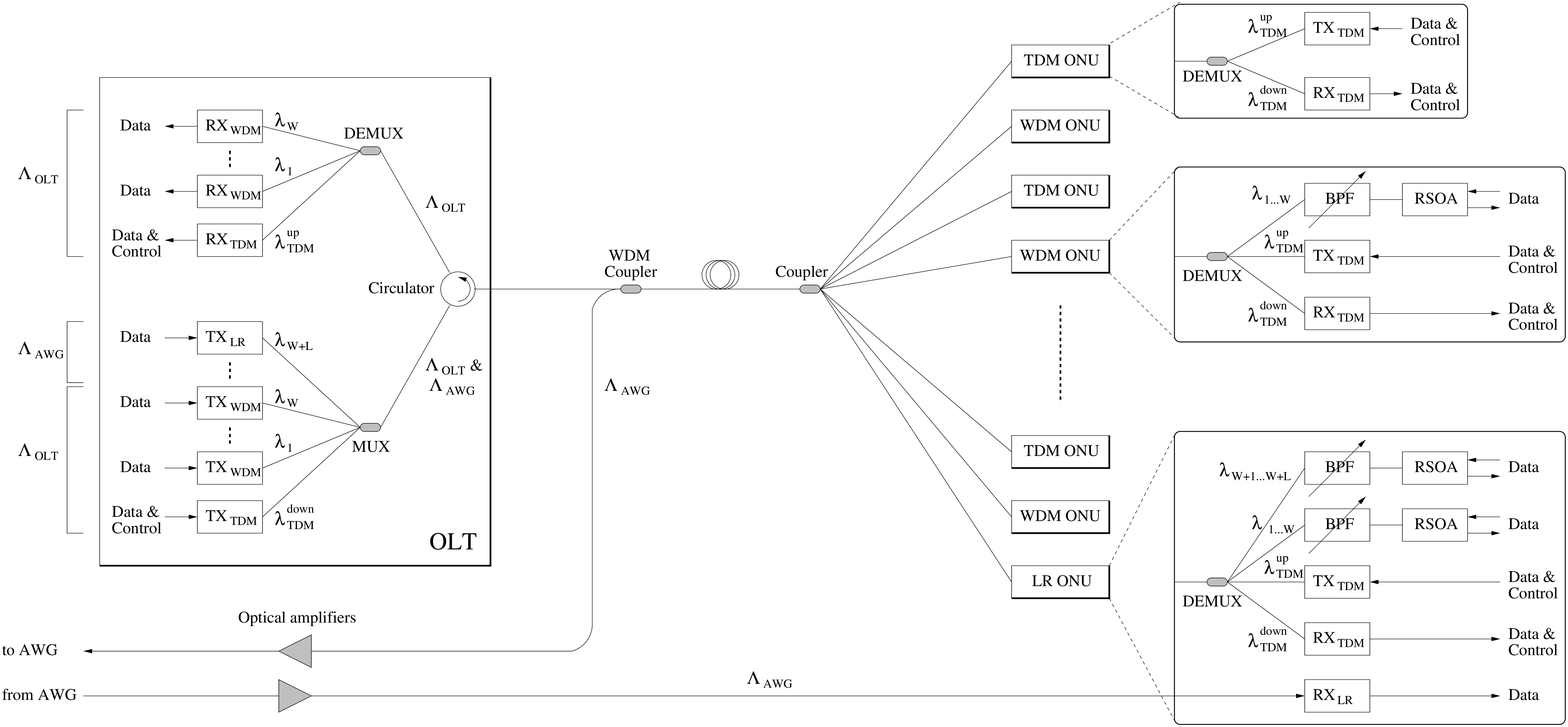}
\caption{Evolutionary upgrade of legacy TDM ONUs to WDM ONUs and
long-reach ONUs (LR ONUs) and their coexistence on the same
NG-PON fiber infrastructure.}
\label{fig:NG-PONs}
\end{center}
\end{figure*}

The MPCP can be extended to support
a wide range of possible WDM ONU structures by exploiting the
reserved bits of GATE, REPORT, and other MPCP
messages~\cite{McMR06}. Similar WDM extensions can be designed for
GPON by exploiting the reserved bits in BWmap, DBRu, and other fields
of each time frame.
In WDM PONs, so-called colorless (i.e.,
wavelength-independent) ONUs should be deployed such that only a
single type of WDM ONU is required, thereby greatly simplifying
inventory, maintenance, and installation~\cite{Effe07}. A promising
approach toward realizing low-cost colorless ONUs is to use a
reflective semiconductor optical amplifier (RSOA) at the ONU to
perform remote modulation, amplification, and reflection of an optical seed
signal sent by the OLT.
The optical seed signal can be either ($i$) a
modulated signal carrying downstream data,
or ($ii$) an unmodulated \emph{empty
  carrier}. In the former case, the ONU reuses the
modulated carrier by means of \emph{remodulation} techniques, e.g.,
FSK for downstream and OOK for upstream~\cite{GrEl08}.

\subsection{Long-reach PON}
Long-reach PONs increase the range and split ratio of conventional TDM
and WDM PONs significantly~\cite{DaGR09,LaPC08,SCLGLBP0909,SM0209}.
State-of-the-art
long-reach PONs are able to have a total length of 100 km potentially
supporting 17 power-splitting TDM PONs, each operating at a different
pair of upstream and downstream wavelength channels and serving up to
256 colorless ONUs, translating into a total of 4352 colorless
ONUs~\cite{TaTo06}.
Importantly, such long-reach PON technologies allow for the
integration of optical access and metro
networks, i.e., broaden the functionality of PONs.
This broadened PON functionality
offers major cost savings by reducing
the number of required optical-electrical-optical (OEO) conversions,
at the expense of optical amplifiers required to compensate for
propagation and splitting losses~\cite{SM0209}.

\subsection{Migration Toward Integrated Access-Metro Networks}
\label{mig:sec}
To provide backward compatibility with legacy infrastructure,
current TDM PONs are expected to evolve
toward NG-PONs in a pay-as-you-grow manner.
Fig.~\ref{fig:NG-PONs}
depicts a tree network architecture for an evolutionary upgrade from
legacy TDM ONUs to WDM ONUs and
long-reach ONUs (LR ONUs), which was originally proposed,
but not formally analyzed, in~\cite{MADM09}.
We briefly review this tree architecture here and incorporate it as a
subnetwork for building an NG-PON in
our original capacity and delay analysis.

\subsubsection{OLT Architecture}
The OLT is equipped with an array of fixed-tuned transmitters (TX) for
transmission of control and data in the downstream direction and a
separate array of fixed-tuned receivers (RX) for reception of control
and data in the upstream direction. Specifically, the OLT deploys one
TX$_{\rm TDM}$ and one RX$_{\rm TDM}$ to send and receive control and
data on the downstream wavelength channel $\lambda_{\rm TDM}^{\rm down}$
and upstream wavelength channel $\lambda_{\rm TDM}^{\rm up}$,
respectively, of the original TDM EPON network (note that
$\lambda_{\rm TDM}^{\rm down}$ and $\lambda_{\rm TDM}^{\rm up}$ are two
different wavelengths). In addition, the OLT may deploy arrays of
fixed-tuned transmitters and receivers for data transmission only (no
control). More precisely, the OLT may deploy $W$ fixed-tuned
transmitters and $W$ fixed-tuned receivers, where $W\geq 0$. These $W$
transmitters and receivers operate on $W$ different wavelength
channels $\lambda_1, \ldots, \lambda_W$ (note that there are exactly
$W$ wavelengths which are alternately used for downstream transmission
and upstream transmission, as described in greater detail in
Section~\ref{sec:op}). The two wavelength channels $\lambda_{\rm
TDM}^{\rm down}$ and $\lambda_{\rm TDM}^{\rm up}$ together with the $W$
wavelength channels make up the waveband $\Lambda_{\rm OLT}$, whose
$2+W$ wavelengths allow for direct optical communication between OLT
and ONUs belonging to the same EPON tree network.

For optical communication between ONUs belonging to different EPON
tree networks, the OLT uses additional $L$ fixed-tuned transmitters
TX$_{\rm LR}$ (but no additional receivers), where $L\geq 0$. These
$L$ transmitters operate on $L$ separate wavelength channels
$\lambda_{W+1},\ldots, \lambda_{W+L}$, which make up the waveband
$\Lambda_{\rm AWG}$. Recall that these $L$ wavelengths do not
necessarily need to be adjacent. These wavelengths are used to
all-optically interconnect different EPON tree networks with suitable
traffic demands (defined shortly).

In the downstream direction, the two wavebands $\Lambda_{\rm OLT}$ and
$\Lambda_{\rm AWG}$ are combined via a multiplexer (MUX) and guided by
the circulator toward the coupler which equally distributes both
wavebands among all $N$ ONUs. In the upstream direction, the WDM
coupler in front of the OLT is used to separate the two wavebands from
each other; the waveband $\Lambda_{\rm OLT}$ is forwarded to the
circulator which guides it onwards to the demultiplexer (DEMUX) that
in turn guides each wavelength channel to a different fixed-tuned
receiver. Whereas the waveband $\Lambda_{\rm AWG}$ is not terminated
at the OLT (therefore no need for receivers at the OLT) and optically
bypasses the OLT, possibly amplified, on its way to the AWG of the
star subnetwork, as explained in Section~\ref{bypass:sec}.

\subsubsection{ONU Architectures}
There are three different types of ONU architecture:
\begin{itemize}
\item{\bf TDM ONU:} The TDM ONU is identical to an ONU of a
conventional TDM EPON network. It is equipped with a single
fixed-tuned transmitter TX$_{\rm TDM}$ and a single fixed-tuned
receiver RX$_{\rm TDM}$ operating on the upstream wavelength channel
$\lambda_{\rm TDM}^{\rm up}$ and downstream wavelength channel
$\lambda_{\rm TDM}^{\rm down}$, respectively. Each wavelength channel in
either direction is used to send both data and control. The DEMUX is
used to separate the two wavelength channels.
\item{\bf WDM ONU:} The WDM ONU is more involved than the TDM ONU in
that it additionally allows to send and receive data on any other
wavelength channel $\lambda_1,\ldots,\lambda_W$ of waveband
$\Lambda_{\rm OLT}$. To do so, the WDM ONU deploys an extra bandpass
filter (BPF) and reflective semiconductor optical amplifier
(RSOA). The BPF can be tuned to any of the wavelengths
$\lambda_1,\ldots,\lambda_W$ and blocks all but one wavelength
$\lambda_i$, $i=1,\ldots, W$. The wavelength $\lambda_i$ passing the
BPF is forwarded to the RSOA. At any given time, the RSOA can be in
either one of the following two operation modes: $(i)$ reception of
downstream data on $\lambda_i$ coming from the OLT, or $(ii)$
transmission of upstream data on $\lambda_i$ originating from the WDM
ONU, as explained in greater detail in Section~\ref{sec:RSOA}.
\item{\bf LR ONU:} The LR ONU builds on the WDM ONU. Similar to the WDM
ONU, the LR ONU has a BPF that is tunable over the wavelengths
$\lambda_1,\ldots,\lambda_W$. In addition, the LR ONU has a RSOA and
BPF tunable over wavelengths $\lambda_{W+1},\ldots,\lambda_{W+L}$,
i.e., this second BPF is tunable over the waveband $\Lambda_{\rm
AWG}$. As a result, the LR ONU can send and receive data also on any
wavelength channel of waveband $\Lambda_{\rm AWG}$, which optically
bypasses the OLT and allows for direct optical communication with
ONUs residing in different EPON tree networks.
Apart from two BPFs
and two RSOAs, the LR ONU deploys an additional multiwavelength receiver
RX$_{\rm LR}$ that is able to simultaneously receive data on all $L$
wavelength channels of waveband $\Lambda_{\rm AWG}$ coming from the
AWG of the star subnetwork, possibly amplified.
\end{itemize}

Note that an EPON tree network may accommodate any combination of
$N_T$ TDM ONUs, $N_W$ WDM ONUs, and $N_L$ LR ONUs, whereby $0\leq N_T,
N_W, N_L\leq N$ and $N_T+N_W+N_L=N$.

\begin{figure}[t]
\begin{center}
\includegraphics[width=.5\textwidth]{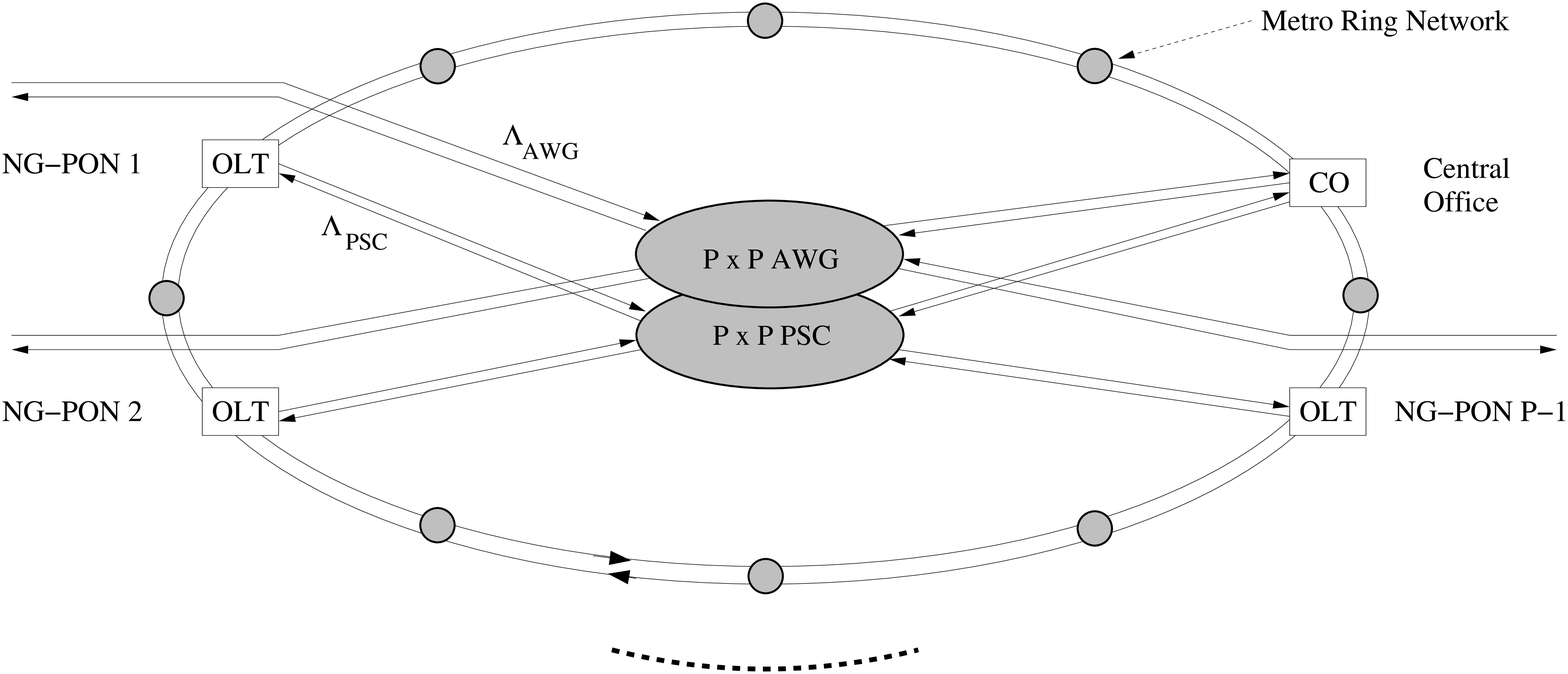}
\caption{Integration of NG-PONs and remote central office using the
  AWG as a wavelength router and a bidirectional metro ring and star
  subnetwork for enhanced resilience.}
\label{fig:AWG}
\end{center}
\end{figure}

As illustrated in Fig.~\ref{fig:AWG},
the OLTs of $P-1,\ P> 2$, NG-PON access networks
(shown in Fig.~\ref{fig:NG-PONs}) and $H$ remote COs in a metro area
may be interconnected with any combination of
$(i)$ a bidirectional metro ring network, e.g., Gigabit Ethernet or
RPR, with $N_r$ ring nodes,
$(ii)$ a wavelength-broadcasting $P\times P$ passive star
coupler (PSC) with $\Lambda_{\rm PSC}$ wavelength channels,
comprising $P$ data wavelength
channels (one assigned to each OLT and CO) and
one control wavelength for conducting a reservation based medium access
control,
and $(iii)$ an AWG employed as a $P\times P$ wavelength router.
Importantly, the AWG
provides any-to-any optical single-hop
connections among the $P-1$ NG-PONs and the CO,
i.e., integrates access and metro networks into one single-hop optical network,
and is therefore the key to vastly improved performance
as demonstrated in Section~\ref{num:sec}.
In addition, the AWG allows for spatial
reuse of all $\Lambda_{\rm AWG}$ wavelengths at each AWG port,
resulting in a significantly increased capacity.
Using two or more of these interconnections options in parallel
enhances network resilience.
In the following, the interconnection network segments are
described at length in separate subsections.

\subsection{Metro Ring Network}
\label{sec:RPR}
The metro ring network, e.g., Gigabit Ethernet or RPR,
interconnects multiple EPON tree networks among
each other as well as to the Internet and server farms. The ring
network consists of $P$ central offices (COs) and $N_r$ ring nodes
equally spaced on the ring ($P=4$ and $N_r=8$ in
Fig.~\ref{fig:AWG}). The CO in the upper right corner of
the figure is assumed to be attached to the Internet and a number of
servers via a common router. We refer to this $H = 1$ CO as the \emph{hotspot}
CO. Each of the other $P - H = 3$ COs is collocated with
the optical line terminal
(OLT) that belongs to the attached EPON tree network.

In case of an RPR metro ring network, the ring is a
optical dual-fiber bidirectional ring network, where each
fiber carries a single wavelength channel (i.e., no wavelength
division multiplexing [WDM]). Each RPR ring node and CO is equipped
with two pairs of fixed-tuned transmitter and fixed-tuned receiver,
one for each fiber ring. Each ring node and CO has separate
(electrical) transit and station queues for either ring. Specifically,
for each ring, a node has one transit queue for in-transit traffic,
one transmission queue for locally generated data packets, and one
reception queue for packets destined for the local node.
In-transit
ring traffic is given priority over station traffic so that in-transit
packets are not lost due to buffer overflow. Thus, the transit path is
lossless, i.e., a packet put on the ring is not dropped at downstream
nodes. On the downside, however, a node ready to send data has to wait
for the transit path to be empty before it can send data. Nodes
perform destination stripping, i.e., the destination node of a given
packet takes the packet off the ring, to let nodes downstream from the
destination node use the ring for packet transmission. Typically,
nodes deploy shortest path routing, i.e., a source node selects the
fiber ring that provides the shortest path to the destination node in
terms of number of traversed intermediate nodes (hops).

\subsection{Star Subnetwork}
\label{StarSubnet:sec}
The $P$ COs of the RPR ring network are interconnected via a
single-hop WDM star subnetwork whose hub is based on a $P\times P$
passive star coupler (PSC) in parallel with a $P\times P$ arrayed
waveguide grating (AWG). The PSC is a wavelength-broadcasting device,
i.e., each wavelength arriving on any PSC input port is equally
distributed to all PSC output ports. In contrast, the AWG is a
wavelength-routing device. The AWG allows for the spatial reuse of all
wavelength channels at each AWG port.
Fig.~\ref{fig:AWG} illustrates spatial wavelength reuse for an
$8\times 8$ AWG ($P=8$) and eight wavelengths $\lambda_1, \ldots,
\lambda_8$. Observe that the same wavelength channel can be
simultaneously deployed at two (and more) AWG input ports without
resulting in channel collisions at the AWG output ports. For instance,
wavelength $\lambda_4$ incident on input ports 1 and 2 is routed to
different output ports 4 and 5, respectively. The wavelength-routing
characteristics of the AWG have the following two implications:
\begin{itemize}
\item Due to the fact that the AWG routes wavelengths arriving at a
given input port independently from all other AWG input ports, no
network-wide scheduling but only local scheduling at each AWG input
port is necessary to avoid channel collisions on the AWG.
\item Note that in Fig.~\ref{fig:AWG} each AWG input port reaches a
given AWG output port on a different wavelength channel. Consequently,
with full spatial wavelength reuse, eight different wavelengths
arrive at each AWG output port simultaneously. To avoid receiver
collisions, each AWG output port must be equipped with a receiver
operating on all eight wavelengths. (A receiver collision occurs if
none of the destination node's receivers is tuned to the wavelength on
which data arrives.)
\end{itemize}
\begin{figure}[t]
\begin{center}
\includegraphics[width=.5\textwidth]{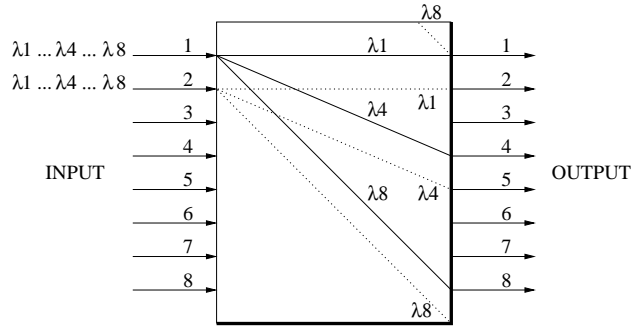}
\caption{Wavelength routing of an $8\times 8$ AWG.}
\label{fig:AWGwlroute}
\end{center}
\end{figure}

The one-way end-to-end propagation delay of the star subnetwork equals
$\tau_{\rm star}$. Each CO is attached to a separate input/output port
of the $P\times P$ AWG and $P\times P$ PSC by means of two pairs of
counterdirectional fiber links. Each fiber going to and coming from
the AWG carries $\Lambda_{\rm AWG}=L$ wavelength channels, whereby $L$
denotes an integer number of wavelengths that do not necessarily need
to be adjacent. Each fiber going to and coming from the PSC carries
$\Lambda_{\rm PSC}=1+h+(P-1)$ wavelength channels, consisting of one
control channel $\lambda_c$, $1\leq h \leq P-1$ dedicated home channels
for the hotspot CO, and $(P-1)$ dedicated home channels, one for each
of the remaining $(P-1)$ COs. The home channels are fixed assigned to
the respective COs for data reception. Data destined for a certain CO
is sent on its corresponding home channel(s).

Each CO has one transmitter and one receiver fixed tuned to the
control channel $\lambda_c$.
In addition, for data reception, each CO (except the
hotspot CO) has a single receiver fixed-tuned to its PSC home
channel. The hotspot CO is equipped with $1\leq h\leq P-1$ receivers
fixed-tuned to its $h$ PSC home channels. For data transmission on the
PSC, each CO (except the hotspot CO) deploys a single transmitter that
can be tuned over the $(P-1)+h$ home channels of the COs. The hotspot
CO deploys $h$ tunable transmitters whose tuning range covers the home
channels of the remaining $(P-1)$ COs as well as the $\Lambda_{\rm
AWG}$ wavelength channels. Thus, among all COs only the hotspot CO is
able to send data on the $\Lambda_{\rm AWG}$ wavelength
channels. Unlike the remaining COs, the hotspot CO is equipped with an
additional multiwavelength receiver operating on all $\Lambda_{\rm
AWG}$ wavelength channels. Hence, only the hotspot CO is able to
receive data on the $\Lambda_{\rm AWG}$ wavelength channels. Note that
the remaining COs are unable to access the $\Lambda_{\rm AWG}$
wavelength channels, which optically bypass these COs and their
collocated OLTs.

\subsubsection{Optical Bypassing}
\label{bypass:sec}
\begin{figure}[t]
\begin{center}
\includegraphics[width=.75\textwidth]{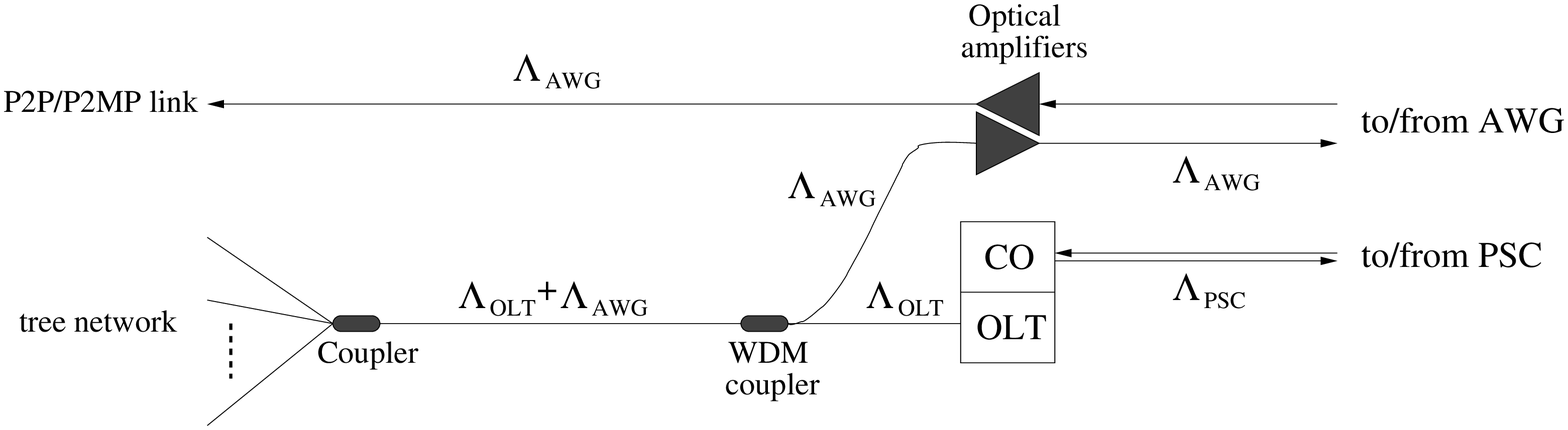}
\caption{Optical bypassing of collocated OLT and CO.}
\label{fig:bypass}
\end{center}
\end{figure}
In the following, we describe how the $\Lambda_{\rm AWG}$ wavelength
channels optically bypass the OLT of Fig.~\ref{fig:NG-PONs} as well as
the collocated CO of Fig.~\ref{fig:AWG}. Fig.~\ref{fig:bypass}
depicts the interconnection of a given EPON tree network and the star
subnetwork, illustrating the optical bypassing of the collocated OLT
and CO. Note that the $\Lambda_{\rm AWG}$ wavelength channels are
carried on the tree network only in the upstream direction, while in
the downstream direction they are carried on a separate point-to-point
(p2p) or point-to-multipoint (p2mp) fiber link. A p2p fiber link
connects a single LR ONU (as illustrated in Fig.~\ref{fig:NG-PONs}) to
the AWG of the star subnetwork, or more generally, a p2mp fiber link
connects multiple LR ONUs (as in Fig.~\ref{fig:AWG}) to the AWG
of the star subnetwork. As shown in Fig.~\ref{fig:bypass}, a WDM
coupler is used on the tree network in front of the OLT to separate
the $\Lambda_{\rm AWG}$ wavelength channels from the $\Lambda_{\rm
OLT}$ wavelength channels and to guide
the $\Lambda_{\rm AWG}$ wavelength channels
directly onward to the AWG
of the star subnetwork, optically amplified if necessary. In doing so,
the $\Lambda_{\rm AWG}$ wavelength channels are able to optically
bypass the CO and OLT without being electrically
terminated. Similarly, the $\Lambda_{\rm AWG}$ wavelength channels
coming from the AWG optically bypass both CO and OLT and directly
travel on the p2p or p2mp link onward to the subset of attached single
LR ONU or multiple LR ONUs, respectively. As a result, the LR ONU(s)
as well as the hotspot CO that send and receive data on any of the
$\Lambda_{\rm AWG}$ wavelength channels are able to communicate
all-optically with each other in a single hop across the AWG of the
star subnetwork.

\section{Operation}
\label{sec:op} In this section, we outline how
the interconnection of the NG PONs across the metro area works.
We first note that each CO performs store-and-forward
transmission with OEO conversion for each packet traversing the CO,
i.e., the packet must first be completely received before it can be
transmitted onto the next subnetwork. Each ring node
performs cut-through transmission for each packet traversing the
node, i.e., the packet is received and forwarded on (ideally) a
bit-by-bit basis.

Before explaining the details of the network operation, we briefly
describe the operation of the RSOA used in the WDM ONU and LR ONU of
Fig.~\ref{fig:NG-PONs}, followed by reporting bandwidth requirements
of ONUs, granting transmission windows to ONUs, dynamic bandwidth
allocation, access control on ring and PSC, as well as considered
traffic types.

\subsection{RSOA}
\label{sec:RSOA}
\begin{figure}[t]
\begin{center}
\includegraphics[width=.95\textwidth]{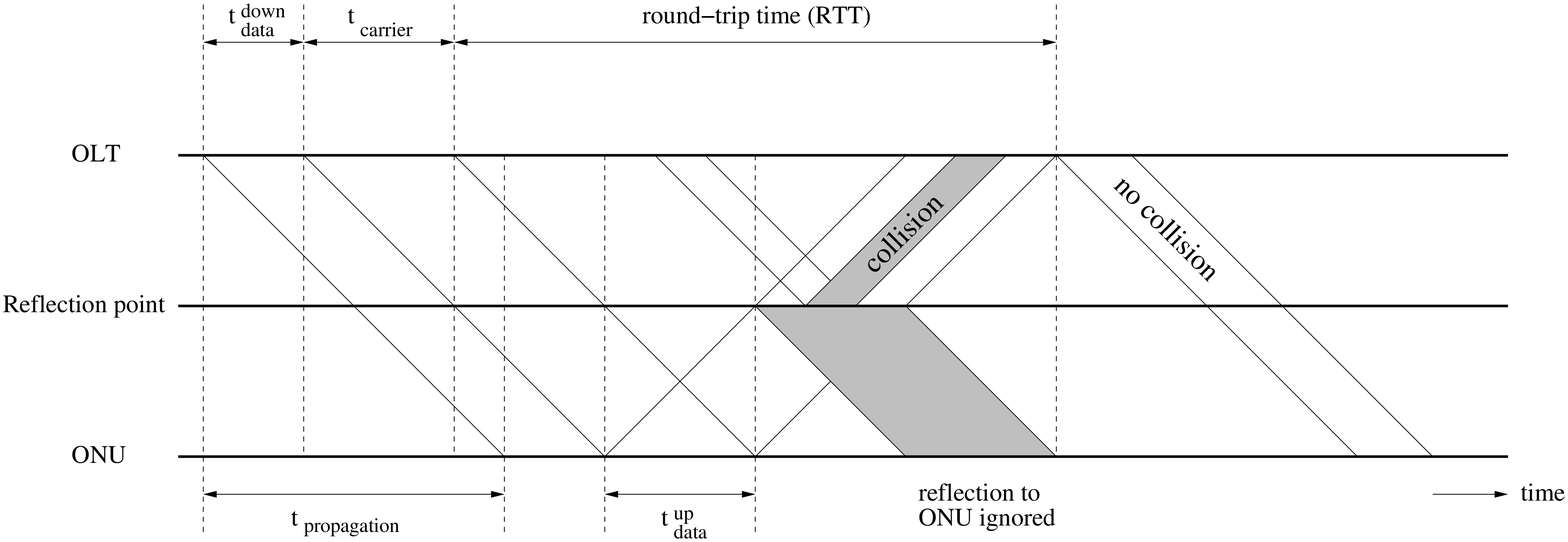}
\caption{RSOA operation modes subject to reflection.}
\label{fig:reflection}
\end{center}
\end{figure}
The RSOA has two different operating modes: $(i)$ reception of
downstream data sent by the OLT to the ONU, and $(ii)$ transmission of
upstream data generated by the ONU and destined for the OLT
(on wavelengths $\lambda_1, \ldots, \lambda_W$),
or for an LR ONU residing in a
different EPON tree network
(on wavelengths $\lambda_{W+1},\ldots, \lambda_L$). The
two modes are alternated on a time basis and at any given time the
RSOA can be in either one of these two modes or in idle state (to be
discussed shortly).

Fig.~\ref{fig:reflection} illustrates the two operation modes $(i)$
and $(ii)$ of an RSOA and also explains the impact of reflection on
the upstream data reception. Suppose that the OLT first sends
downstream data during the time interval $t_{data}^{\rm down}$ using one
of its fixed-tuned transmitters TX$_{\rm WDM}$ or TX$_{\rm LR}$. After
the propagation delay $t_{propagation}$, or briefly $\tau_T$, the data
arrives at the destination ONU.

The upstream data transmission from ONU to OLT is a bit more
involved due to the fact that the RSOA does not have its own light
source. As a consequence, the OLT has to generate light by using one
of its WDM transmitters TX$_{\rm WDM}$ (in case of a WDM ONU or LR
ONU) or one of its LR transmitters TX$_{\rm LR}$ (only in case of an
LR ONU) and send it downstream to the ONU. We consider two
approaches for supplying light to the RSOA: the empty carrier
approach, and the reflection of downstream data signal approach. We
first describe the empty carrier approach where a light carrier
signal that does not carry any data is supplied to the RSOA for
upstream data transmission. In Fig.~\ref{fig:reflection},
suppose that after transmitting its downstream data, the OLT sends
the generated light to the ONU during the time period $t_{carrier}$,
or briefly $t_c$, which can be of any arbitrary length. Note that
this light does not convey any downstream data; it may be viewed as
an ``empty'' carrier arriving after $t_p$ at the ONU. The RSOA acts
as a mirror and the ONU uses the carrier light reflected by the RSOA
for sending its upstream data to the OLT during the time interval
$t_{data}^{\rm up}$. The upstream data transmission takes $t_p$ to
arrive at the OLT. Now, to guarantee that the upstream data is
received by the OLT without collision, the OLT must not use the same
wavelength for downstream data transmission until the upstream data
is completely received by the OLT. In other words, after generating
the light and sending it downstream to the ONU, the OLT has to wait
for one round-trip time (RTT), i.e., $2 \tau_T$, until it is allowed
to use the same wavelength again for downstream data transmission.

To better understand this
constraint, Fig.~\ref{fig:reflection} illustrates the case where the
OLT does not wait for one RTT and starts its next downstream data
transmission before one RTT has elapsed. The second downstream data
transmission might be reflected at a reflection point (e.g., splice or
connector) somewhere between the OLT and ONU and interfere with the
upstream data transmission of the ONU, resulting in a
collision. Clearly, by deferring its next downstream data transmission
by at least one RTT, the OLT can avoid any collisions. However, note
that while waiting for the wavelength to become available, the OLT may
use the downstream wavelength channel $\lambda_{\rm TDM}^{\rm down}$
and the other WDM wavelength channels to
send data to any (WDM, LR, and TDM) ONU. Also note that this
restriction holds only for data but not for carrier sent in the
downstream direction. For instance, in Fig.~\ref{fig:reflection} the
OLT might generate a second carrier on the same wavelength destined
for a different ONU right after $t_c$, provided that the upstream data
transmission of the second ONU does not overlap with the first one due
to different propagation delays between OLT and the two ONUs.
With the reflection of downstream data signal approach,
the downstream data signal is used as carrier signal for upstream data
transmission.

\subsection{REPORT Message}
Each ONU monitors traffic incoming from its attached users and
classify it into one or more different traffic types. Specifically,
each TDM ONU and each WDM ONU considers all incoming traffic the same
and store it in a single first-in-first-out (FIFO) queue (for service
differentiation we might later consider multiple FIFO queues, one for
each traffic class). In each polling cycle, every TDM/WDM ONU sends
its current bandwidth requirements (queue occupancy) in its assigned
REPORT message to the OLT on wavelength channel $\lambda_{\rm
TDM}^{\rm up}$. In contrast, each LR ONU splits incoming traffic into two
different traffic types, using two separate FIFO queues. The first
queue, which we call \emph{OLT queue}, stores traffic to be sent
on any of the $\Lambda_{\rm OLT}$ upstream wavelengths. The second
queue, which we call \emph{AWG queue},
stores traffic to be sent on one of the $\Lambda_{\rm AWG}$ wavelength channels
(as discussed in more detail shortly). In each polling cycle,
every LR ONU reports its current bandwidth requirements (queue
occupancies) on both $\Lambda_{\rm OLT}$ and $\Lambda_{\rm AWG}$ in a
single REPORT message to the OLT on $\lambda_{\rm TDM}^{\rm up}$. More
precisely, an LR ONU writes its bandwidth requirements on
$\Lambda_{\rm OLT}$ into the REPORT message, similar to a TDM or WDM
ONU. In addition, an LR ONU writes its bandwidth requirements on
$\Lambda_{\rm AWG}$ and the addresses of the corresponding destination
ONUs into the REPORT message. Thus, the bandwidth requirements on
$\Lambda_{\rm AWG}$ (including destination ONU address) ride piggyback
on those on $\Lambda_{\rm OLT}$ within the same REPORT message.

\subsection{GATE Message}
After receiving the REPORT message(s), the OLT sends a GATE message to
each ONU on the downstream wavelength channel $\lambda_{\rm
TDM}^{\rm down}$. The GATE message contains one or two transmission
grants, depending on the ONU type. Specifically, for a TDM ONU the
GATE message contains a single grant that specifies the start time and
duration of its allocated upstream TDM transmission window on
wavelength channel $\lambda_{\rm TDM}^{\rm up}$. The TDM transmission
window on $\lambda_{\rm TDM}^{\rm up}$ is also used by the TDM ONU to send
its next REPORT message to the OLT. Conversely, for WDM ONUs and LR
ONUs the GATE message contains two transmission grants, the TDM
transmission window for $\lambda_{\rm TDM}^{\rm up}$ and another BPF
window for a wavelength channel within $\Lambda_{\rm OLT}$ or
$\Lambda_{\rm AWG}$. More precisely, for a WDM ONU the first
transmission grant specifies the allocated TDM transmission window on
$\lambda_{\rm TDM}^{\rm up}$ used by the WDM ONU to send data, if any, and
the next REPORT message to the OLT. The second transmission grant
specifies $(i)$ the ``best'' wavelength channel within $\Lambda_{\rm
OLT}$, and $(ii)$ the start time and duration of the BPF window during
which the WDM ONU has to tune its BPF to the ``best'' wavelength for
reception of data coming from the OLT and transmission of data going
to the OLT.
Note that the two windows assigned to a given WDM ONU are
allowed to overlap in time (partly or fully in any arbitrary way).
LR ONUs are handled the same way as WDM ONUs, except that the OLT can
assign an additional BPF window on $\Lambda_{\rm AWG}$ to a given source LR ONU
for transmission of eligible traffic if the destination ONU is an LR
ONU (which is known by the OLT through the piggybacked destination ONU
address in the respective REPORT message). In case of a destination LR
ONU, the second transmission grant in the GATE message specifies $(i)$
the appropriate wavelength channel within $\Lambda_{\rm AWG}$ that
interconnects the source LR ONU with the destination LR ONU across the
AWG and $(ii)$ the start time and duration of the BPF window during
which the source LR ONU has to tune its BPF to the selected wavelength
for reception of data coming from the OLT and transmission of data
going to the destination LR ONU, thereby optically bypassing the
OLT. Otherwise, if the destination ONU is not an LR ONU, the source LR
ONU is handled similar to a WDM ONU, as discussed above.

\subsection{Dynamic Bandwidth Allocation}
The OLT dynamically allocates bandwidth on $\Lambda_{\rm OLT}$ and
$\Lambda_{\rm AWG}$ to its attached ONUs using the \emph{gated}
service discipline, i.e., the OLT does not impose any limit on the
granted transmission windows and assigns each ONU as much bandwidth as
requested in its REPORT message.

\subsection{Access Control on Ring and PSC}
RPR ring nodes as well as ONUs unable to send and receive data
across the AWG, send their data on the tree, ring, and/or PSC
(typically along the shortest path in terms of hops).
Channel
access on the dual-fiber ring is governed by RPR protocols,
described in Section~\ref{sec:RPR}. On the PSC, time is divided into
periodically recurring frames. Each frame on control channel
$\lambda_c$ consists of $P$ control slots, each dedicated to a
different CO. Each CO stores data packets to be forwarded on the
PSC. For each stored data packet the CO broadcasts a control packet
on the PSC to all COs in its assigned control slot. A control packet
consists of two fields: $(i)$ destination address and $(ii)$ length
of the corresponding data packet. After $\tau_{\rm star}$, all COs
receive the control packet and build a common distributed
transmission schedule for collision-free transmission of the
corresponding data packet on the home channel(s) of the destination
CO at the earliest possible time. The destination CO forwards the
received data packet toward the final destination node.

\section{Capacity Analysis}
\label{analysis:sec}

\subsection{Network Model}

Let $C_T^k$, $C_W^k$, $C_A$, $C_P$, and $C_{RPR}$ denote
the transmission capacities [in bits
per second] of one TDM, WDM, AWG, PSC, and RPR channel,
respectively, whereby $k$ indexes the central office (CO) that the
TDM or WDM channel connects to.
We assume that both TDM upstream and downstream channel
at CO $k$ have each a capacity of $C_T^k$.
We denote $W^k$ for the number of WDM channels connecting to
CO $k$, whereby each channel has capacity $C_W^k$.
We define these capacities as ``payload'' transmission capacities
in the sense that they already account for overheads that
are proportional to the
transmitted payload, such as the Preamble and Inter Packet Gap.
The fixed overheads per grant, such as MPCP Report and
guard times, are not considered.

We denote $\mathcal{N}$ for the set of nodes that act as (payload)
traffic sources and destinations. Specifically, $\mathcal{N}$
contains all ONUs in the tree subnetworks, the RPR ring nodes, and
the hotspot nodes (whereby there may be none, one, or several
hotspots). The COs, except for the hotspots, do not locally generate
traffic. We denote  $\mathcal{C}_k$ for the set of all ONUs that are
connected to CO $k$. Additionally, we denote $\mathcal{C}^T_k$ for
the set of TDM ONUs connected to CO $k$, and analogously
$\mathcal{C}^W_k$ for the set of WDM ONUs connected to CO $k$, and
$\mathcal{C}^L_k$ for the set of LR ONUs attached to ONU $k$.

We model the packet generation rates by a traffic matrix
$T=(T(i,j)),\ i, j \in \mathcal{N}$,
where $T(i,j)$ represents the number of packets per second
that are generated at a node $i$ and are destined to node $j$.
Note that for a network with
$(P-H)\cdot N$ ONUs, $N_r$ ring nodes, and $H$ hotspot
COs the traffic matrix has
$((P-H)\cdot N + N_r + H)\times ((P-H)\cdot N +N_r + H)$ elements.
For the stability analysis we assume that the traffic generation is
ergodic and stationary.
Importantly, the traffic matrix $T=(T(i,j)),\ i, j \in \mathcal{N}$,
accounts only for the traffic that is \textit{not} sent over the AWG.
Since the operation of the AWG can be analyzed separately from the
rest of the network, we consider AWG traffic separately
in Section~\ref{AWGintensities:sec}.
A natural assumption could be to assume that traffic is equally
distributed from all ONUs to all other ONUs and/or RPR ring nodes,
except possibly the hotspot(s). But we do not need to restrict our
analysis to this special uniform traffic case.

We suppose that there is a packet length distribution (in
bits) with mean $\bar{L}$ and variance $\sigma_L^2$.
For notational simplicity we suppose that this
distribution is the same for all source nodes.
Note that, on
average, node $i$ sends $\bar{L} \cdot T(i,j)$ bits per second
to node $j$.

\subsubsection{Traffic Rates in Ring/PSC Star Subnetwork}
First let us consider the traffic that is eventually sent over the
ring/PSC star subnetwork.
The traffic that arrives from the ONUs at CO $k$ (over the
conventional TDM or WDM channels) and is destined to another CO
$l,\ l\neq k$, enters the ring/PSC star subnetwork.
Hence, the packet rate of the ring/PSC star traffic between the
two COs $k$ and $l$,\ $k\neq l$, is given by
\begin{equation}
\sigma(k,l):=\sum_{i \in \mathcal{C}_k,\ j \in \mathcal{C}_l}
T(i,j).
\end{equation}
Note that traffic from an ONU $i$ to another ONU $j,\ j \neq i$,
attached to the same CO $k$, i.e., with $i,j \in \mathcal{C}_k$,
does not enter the ring/PSC star subnetwork.

For an RPR ring node or a
hotspot $k$ and a CO $l$ we define
\begin{equation}
\sigma(k,l) := \sum_{j \in \mathcal{C}_l} T(k,j),
\end{equation}
and
for a CO $k$ and an RPR ring node or a hotspot $l$
\begin{equation}
\sigma(k,l):=\sum_{i \in \mathcal{C}_k} T(i,l).
\end{equation}
Finally, for traffic from an RPR ring node/hotspot $k$
to another RPR ring node/hotspot $l$, we have
\begin{equation}
\sigma(k,l):=T(k,l).
\end{equation}
Note that the defined $\sigma(k,l)$ [in packets/second]
completely determine the
packet traffic rates in the
ring/PSC star subnetwork.

For each node $i,\ i \in \mathcal{N}$, we denote
\begin{equation} \label{sigma_i:defn}
\sigma(i) := \sum_{l \in \mathcal{N}} T(i,l),
\end{equation}
for the total packet traffic generation
rate (in packets/second)
at node $i$.
Note that this packet rate only accounts for the
traffic that is \textit{not} sent over the AWG.

\subsubsection{Traffic Rates in AWG Star Subnetwork}
\label{AWGintensities:sec}
In addition to traffic that is sent over the tree and
ring/PSC star subnetwork, an LR ONU or hotspot may generate
traffic that is eligible for transmission over the AWG
via the p2p/p2mp links.
We suppose that an LR
ONU/hotspot that generates a packet for another LR ONU/hotspot
that {\it can} be
reached via the AWG, i.e., there exists a
p2p/p2mp link between the considered LR ONUs/hotspots,
{\it will} send that packet
over the AWG, and {\it not} over tree and ring/PSC star subnetworks.
Formally, we let
$c(k,l)$ denote the number of wavelength channels
on the p2p/p2mp link
 between CO $k$ and CO $l$.
We denote $T^A(i,j)$ for the packet traffic rate
in number of generated packets per second from LR ONU/hotspot
$i$ to LR ONU/hotspot $j$.
If $c(k,l)=0$, i.e., if there is no
p2p/p2mp link from LR ONU $i \in \mathcal{C}^L_k$
to LR ONU $j \in \mathcal{C}^L_l$, then $T^A(i,j) = 0$.
For LR ONU/hotspot $i$, we define
\begin{equation}
\sigma^A(i) := \sum_{l \in \mathcal{N}}
        T^A(i,l),
\end{equation}
as the total packet generation rate [in packets/second]
of traffic transmitted over the AWG.
Note that an LR ONU/hotspot $i$ may also generate traffic
that is transmitted over the tree and ring/PSC star subnetwork;
specifically, for RPR ring node, TDM ONU, and WDM ONU destinations, or
for LR ONU/hotspot destinations not reachable from LR ONU/hotspot
$i$ via the AWG.
This ``non-AWG'' traffic is accounted for in $\sigma(i)$
given by (\ref{sigma_i:defn}).

\subsection{Capacity of EPON Tree Subnetwork}
\label{treecap:sec}
\subsubsection{Upstream Capacity}
Each ONU must not generate more traffic than it can send in
the long term average.
A TDM ONU can only transmit on the upstream TDM channel, whereas
a WDM or LR ONU can transmit on the upstream TDM channel
and one WDM channel.
\begin{equation}
\bar{L} \cdot \sigma(i) <
\begin{cases}
C_T^k & i \in \mathcal{C}^T_k\\ C_W^k + C_T^k &
i \in \mathcal{C}^W_k \cup \mathcal{C}^L_k
\end{cases}
\end{equation}
Similarly, we require for
each LR ONU that
\begin{equation}
    \bar{L} \cdot \sigma^A(i) < C_A \label{eqntbc}.
\end{equation}


All TDM ONUs at a given CO $k$ considered together must not transmit more than
$C_T^k$ on the upstream TDM channel
\begin{equation}
\lambda^{T,u,k} :=
       \sum_{i \in \mathcal{C}^T_k} \bar{L} \cdot \sigma(i)
         < C_T^k. \label{eqn:tdmcap}
\end{equation}

The WDM and LR ONUs can send over the WDM channels
and they can use the remaining bandwidth of the upstream TDM channel:
\begin{equation}
      \lambda^{W,u,k}
            := \sum_{i \in \mathcal{C}^W_k \cup \mathcal{C}^L_k} \bar{L}
                \cdot \sigma(i)
            < W^k C_W^k + C_T^k - \lambda^{T,u,k}. \label{eqn:wdmcap}
\end{equation}

We note that the RSOA operation does not reduce the capacity. The
finite (bounded) switchover time from one transmission direction to
the other becomes negligible when queues grow long. In particular,
for heavy traffic load and correspondingly very long continuous
upstream or downstream transmissions, the switchover time becomes
negligible. However, the RSOA operation has an impact on the delay,
as analyzed in Section~\ref{delay_WDM:sec}.

\subsubsection{Downstream Capacity}
The traffic arriving for the TDM ONUs has to be accommodated on the
downstream TDM channel:
\begin{equation} \label{lambda_TDM_down:eqn}
      \lambda^{T,d,k}
          := \sum_{j \in \mathcal{C}^T_k}
              \sum_{l\in \mathcal{N}} \bar{L} \cdot T(l,j)
        < C_T^k.
\end{equation}

The WDM channels may be used for upstream or downstream transmission.
We investigate two possibilities.

\paragraph{Reflection of Downstream Data Signal for Upstream Data Transmission}
Suppose the ONUs are equipped with a device that allows to use
the downstream data signal as carrier signal for upstream data transmission.
The OLT would either send downstream data; or, if there is no data to be sent,
it would transmit `empty' carrier.
For this operating scenario, upstream and downstream transmissions
are completely independent.

The restriction for upstream traffic is given in
(\ref{eqn:wdmcap}).
For the downstream traffic we obtain
\begin{equation} \label{lambda_WDM_down:eqn}
   \lambda^{W,d,k}
        := \sum_{j \in \mathcal{C}^W_k \cup \mathcal{C}^L_k }
             \sum_{l \in \mathcal{N}} \bar{L}
             \cdot T(l,j)
        < W^k C_W^k + C_T^k - \lambda^{T,d,k}.
\end{equation}

\paragraph{Empty Carrier for Upstream Data Transmission}
If upstream data transmissions require an empty carrier signal,
we can only use a WDM channel for either upstream or
downstream transmission, resulting in the additional restriction that
\begin{equation}  \label{lambda_WDM_down_empty:eqn}
    \lambda^{W,d,k} + \lambda^{T,d,k} +
             \lambda^{T,u,k} + \lambda^{W,u,k}
          < C_T^k + W^k C_W^k.
\end{equation}

\subsection{Capacity of the Ring/PSC Star Subnetwork}
\label{cap_ringstar:sec}
Note that the $\sigma(k,l)$ defined above correspond to the respective
$\sigma(i,j)$ in \cite{MHSR05}.
Further, we can
introduce and calculate the analogous
probabilities to $p_{k,l}(e)$ and $p_{k,l}(m,n)$ in \cite{MHSR05}.
Specifically, for our context, we introduce the probabilities
$p_{i,j}(k,l)$ that traffic from a node $i \in \mathcal{N}$ to a node
$j \in \mathcal{N}$ traverses the network link $(k,l)$.
These probabilities can be precomputed for given traffic matrices
$T$ and $T^A$, similar to~\cite{MHSR05}, for any pair of nodes in the
network.
The stability condition
of the ring subnet is given by (8) in \cite{MHSR05}.

The mode of operation of the PSC in STARGATE is simpler than in
\cite{MHSR05}, in that there are no collisions. Note that all
traffic that is generated for destination CO $l$ has to queue up in
a {\it virtual queue} that is calculated by all COs. Therefore, for
the capacity evaluations it suffices to require that $\lambda^P(l)$,
the total rate of traffic (in bit/second) going from the PSC
into CO $l$, does not overload the PSC wavelength channel
\begin{equation}  \label{lambdaPl:eqn}
  \lambda^P(l)
      := \sum_{\text{CO $k$, $k\neq l$}}
                \sum_{i,j\in \mathcal{N}} p_{i,j}(k,l) \cdot \bar{L}
                  \cdot \sigma(i,j)
           < h(l) C_P,
\end{equation}
whereby $h(l)$ denotes the number of home channels of CO $l$.

\subsection{Capacity of the AWG}
\label{awgcap:sec}
As noted in Section~\ref{StarSubnet:sec}, the potential for
collisions exists in the AWG star subnetwork only at the AWG input
ports (local scheduling by the CO avoids actual collisions). For the
capacity analysis, it suffices therefore to focus on the wavelength
channels running from the LR ONUs at a given CO to the corresponding
AWG input port. More specifically, the traffic generated by the LR
ONUs of CO $k$ destined toward LR ONUs at CO $l$ must be
accommodated on the wavelength channel(s) that are routed from the
AWG input port of CO $k$ to the AWG output port leading to CO $l$.
Therefore, for analytical purposes, we can treat the AWG operation
as if there were $c(k,l),\ c(k,l) \geq 0$, separate  wavelength
channels from CO $k$ to CO $l,\ l \neq k$. These channels do not
influence each other (in terms of multiple access). Therefore, we
obtain the following restriction for $\lambda^A(k,l)$, the total
average rate of traffic (in bits/second) transmitted from CO (or
hotspot) $k$ over the AWG to CO (or hotspot) $l$:
\begin{equation}  \label{lambdaA:eqn}
    \lambda^A(k,l):=\sum_{i \in \mathcal{C}^L_k,\ j \in \mathcal{C}^L_l}
         \bar{L} \cdot T^A(i,j)
          < C_A \cdot c(k,l).
\end{equation}
If $c(k,l)=0$, i.e., there are no wavelength channels from
CO $k$ to CO $l$ over the AWG, the restriction has to be understood
as $=0$, not $<0$.

\subsection{Summary of Capacity Analysis}
We summarize the capacity analysis by noting that
for given traffic patterns $T(i,l)$ and $T^A(i,l)$ it is relatively
straightforward to obtain from the capacity constraints
in Sections~\ref{treecap:sec}--\ref{awgcap:sec} bounds on the
mean aggregate network throughput.
In particular, we denote
\begin{eqnarray}
r_T = \bar{L} \sum_{i \in \mathcal{N}}
      \left[ \sigma(i) + \sigma^A(i) \right]
\end{eqnarray}
as the total generated traffic
[in bit/second], which is equivalent to the total mean aggregate
throughput of the network.
Each capacity constraint results in an upper bound
on $r_T$. 
The tightest bound identifies the network bottleneck limiting
the mean aggregate throughput, as illustrated in
Section~\ref{intaccmet:sec}.

\section{Delay Analysis}
\label{sec:analysisdel}
For the delay analysis we consider a networking scenario with
many legacy TDM ONUs and a few upgraded WDM and LR ONUs in
the EPON subnetworks.
For such a scenario, the traffic rate on the upstream TDM channel
is typically significantly higher than on the WDM channels:
\begin{equation}
  \sum_{i \in \mathcal{C}^W_k \cup \mathcal{C}^L_k}
     \sigma(i) \ll
   W^k \sum_{i \in \mathcal{C}^T_k} \sigma(i).
\end{equation}
We leave the complementary scenario in which the WDM channels carry
more load than the TDM channel for future work.
In such a scenario the TDM channel becomes
essentially a WDM channel since it will also be used by WDM and LR
ONUs, leading to a situation where one can essentially
neglect the special position of the TDM channel and consider it a WDM channel.

For the considered highly loaded TDM channel scenario, we have
for the downstream TDM channel:
\begin{equation}
       \lambda^{W,d,k} \ll W^k
        \lambda^{T,d,k}.
\end{equation}
In the considered scenario with highly loaded
TDM channels, the WDM and LR ONUs practically do not
transmit or receive payload data on the
TDM channels.
This is in some sense an additional restriction, leading to
a possibly higher delay, i.e., our analysis should lead to an upper
bound for the delay.
Note also that in the considered scenario the delays
for reporting on the upstream TDM channel
and transmitting grants on the downstream TDM channel are
governed by the delays on the TDM channels.

We denote $\tau_T$, $\tau_P$, and $\tau_A$ [in seconds]
for the one-way propagation delay over the EPON tree network,
the PSC star subnetwork, and the AWG star subnetwork, respectively.
For the delay analysis we require that the
traffic that is generated at node $i$ and destined to node $j$
is Poisson with packet generation rate
$T(i,j)$ [packets/second] and independent of the traffic for
all other combinations $i'$, $j'$.
For notational convenience, we define
\begin{eqnarray}
\Phi(\rho) := \frac{\rho
     \left(
        \frac{\sigma_L^2}{\bar{L}} +  \bar{L}
                                              \right)} {2 C (1 - \rho)}
\end{eqnarray} to denote the mean queuing delay
in an M/G/1 queue according to the Pollaczek-Khinchine
formula~\cite{Klein75} as a function of the (relative) load $\rho$ defined
as the traffic rate $\lambda$ [bit/second]
normalized by the channel bit rate $C$ [bit/second]
for the considered packet size mean $\bar{L}$ and variance
$\sigma_L^2$.

\subsection{Delay on the Upstream and Downstream TDM channels}

The long run average traffic rate on the downstream TDM channel
from CO $k$ is
$\lambda^{T,d,k}$ given in (\ref{lambda_TDM_down:eqn}), resulting
in a load $\rho^{T,d,k} = \lambda^{T,d,k} / C_T^k$.
Thus, an initial estimate of the queueing delay of a packet prior to
transmission on the downstream TDM channel is approximately given by
$\Phi( \rho^{T,d,k})$.
This delay does not consider that this traffic has
already traversed preceding nodes feeding into the downstream traffic
at CO $k$. Specifically, the adjacent RPR nodes (via the RPR ring)
and the other COs $l,\ l \neq k$ (via the PSC) supply downstream TDM
traffic to CO $k$.

Applying the approximate method of Bux and Schlatter~\cite{BuS83}
to our setting,
we compensate for the queueing delay at the preceding nodes by subtracting
a correction term $B^{T,d,k}$
from the queueing delay $\Phi( \rho^{T,d,k})$ for the
aggregate downstream traffic $\rho^{T,d,k}$.
Following~\cite{BuS83},
the correction term $B^{T,d,k}$
is the sum of the queueing delays for the individual
traffic stream that flow into CO $k$ from adjacent nodes
\textit{and} leave over the arc of
interest, namely the downstream TDM channel.
In particular,
\begin{equation}
 B^{T,d,k} = \sum_{\text{RPR node $l$ adjacent to CO $k$}} \Phi(\rho^{l,R,T,k})
                + \sum_{\text{CO $l$}}
        \Phi(\rho^{l,P,T,k}). \label{rho:into:T0}
\end{equation}
with
\begin{equation}
 \rho^{l,R,T,k}=\sum_{i\in \mathcal{N}}
       \sum_{j\in\mathcal{C}^T_k}  p_{i,j}(l,k) T(i,j)
           \bar{L}/C_{RPR} \label{rho:into:T1}
\end{equation}
denoting the load due to traffic flowing from the
adjacent RPR node $l$ over the RPR ring to reach one of the TDM ONUs at
CO $k$.
For the evaluation of these traffic loads we utilize the probability
$p_{i,j}(l,k)$ that traffic from a node $i \in \mathcal{N}$ to a node
$j \in \mathcal{N}$ traverses the network link $(l,k)$, as
defined in Section~\ref{cap_ringstar:sec}.
Further, we evaluate
\begin{equation}
 \rho^{l,P,T,k}=\sum_{i\in \mathcal{N}}
   \sum_{j\in\mathcal{C}^T_k}  p_{i,j}(l,k) T(i,j)
           \bar{L}/C_{P} \label{rho:into:T2}
\end{equation}
for the load from a CO $l,\ l \neq k$, over the PSC to a TDM ONU at
CO $k$.

Adding the average transmission delay $\bar{L} / C_T^k$
and the downstream propagation delay $\tau_T$ we obtain the
total delay as approximately
\begin{equation} \label{Dtdk:eqn}
D^{T,d,k,E} :=
  \Phi \left( \rho^{T,d,k} \right) +
    \tau_{E} + \frac{\bar{L}}{C_T^k} - B^{T,d,k},
\end{equation}

The mean of the residual cycle length on the upstream TDM channel
until a generated packet is reported is approximately
$\tau_T$.
In addition, there is a delay of $2 \tau_T$ plus the residual
transmission time of a packet when the downstream channel is
busy, i.e., $\bar{L}/(2C_T) \cdot \rho^{T,d,k}$,
between transmitting a report and receiving the corresponding grant.
Adding queueing delay, packet transmission, and propagation delays
gives
\begin{eqnarray} \label{DTuk:eqn}
D^{T,u,k,E} = 4 \tau_T
            + \Phi(\rho^{T,u,k})+ \frac{\bar{L}}{C_T} +
      \frac{\bar{L}}{2C_T} \rho^{T,d,k}.
\end{eqnarray}

Note that for our NG PON interconnection
network there is an additional delay due
to the queuing of the gate message prior to transmission on the
downstream TDM channel. We consider two approaches for the
downstream transmission of grant messages. (A) Without any priority
for grant messages, the grant message has to queue up with regular
downstream packet traffic, resulting in an additional delay
component of $\Phi( \rho^{T,d,k})$. (B) A grant message could be
given (non-preemptive) priority over data packets as follows. If
there is currently no downstream data packet transmission ongoing,
then immediately send the grant. If there is currently a downstream
data packet transmission ongoing, then transmit the grant when the
current packet transmission is complete. For this priority policy,
the additional delay component is zero when the channel is idle and
the residual transmission time of the packet when the channel is
busy, i.e., $0 \cdot (1- \rho^{T,d,k}) + \bar{L}/(2C^k_T) \cdot \rho^{T,d,k}$.

We remark that the analysis of the upstream TDM cycle length leading to
(\ref{DTuk:eqn}) assumes a delay of
$2 \tau_T$ between sending the report and receipt of the
corresponding grant.
The grant has the additional delay component
(of $\Phi( \rho^{T,d,k})$ or $\rho^{T,d,k} \bar{L}/2$)
due to queueing of the grant prior to being transmitted on
the downstream channel.
Hence, the upstream TDM cycle is longer than reflected in our approximate
analysis.
A more exact analysis of the upstream TDM cycle that captures
the inter-dependencies with the grant transmissions on the
TDM downstream channel is left for future work.

\subsection{Delay on the EPON WDM Channels}
\label{delay_WDM:sec}

\subsubsection{Reflection of Downstream Data Signal for Upstream
          Data Transmission}

When reflecting the downstream data signal for
upstream data transmission, the $W^k$ WDM channels at CO $k$ can be
continuously used for downstream and upstream transmission.
The traffic rate for the downstream WDM channels is given by
$\lambda^{W,d,k}$ (\ref{lambda_WDM_down:eqn}) and there
are $W^k$ channels, each with transmission rate $C_W^k$.
The queueing delay could be approximated by the
queueing delay in an M/G/$W^k$ queueing system.
Since there is no explicit delay formula for such a system,
we further approximate
the delay by considering an M/G/1 queue with a server with
transmission rate $W^k C_W^k$, i.e., we consider an
M/G/1 queue with load
$\rho^{W,d,k} = \lambda^{W,d,k} / (W^k C_W^k) $.
We obtain the total delay as approximately,
\begin{eqnarray}
 D^{W,d,k,E} \approx
 \Phi \left( \frac{\lambda^{W,d,k}}{W^k C_W^k} \right) + \tau_T
      +\frac{\bar{L}}{C_T^k} - B^{W,d,k},
\end{eqnarray}
where $B^{W,d,k}$ is defined analogously to (\ref{rho:into:T0}),
(\ref{rho:into:T1}), and (\ref{rho:into:T2}) with $\mathcal{C}_T^k$
replaced by $\mathcal{C}_W^k\cup \mathcal{C}_L^k$.

The upstream data transmissions have additional delay components due to
the reporting and granting procedure:
\begin{itemize}
\item The residual time of the upstream TDM channel
$\tau_T$
to account for the delay
from packet generation
until transmission of the corresponding report.
\item the round trip propagation delay $2 \tau_T$ to account for the
upstream propagation of the report and downstream propagation of the
grant.
\item the queueing delay for the grant prior to its
transmission on the downstream TDM channel, which is
$\Phi( \rho^{T,d,k})$ without priority for grants,
and $\rho^{T,d,k}\bar{L}/(2C)$ with priority for the grant messages
(considered in (\ref{DWuk:eqn}).
\end{itemize}

Thus, we obtain with the upstream traffic rate
$\lambda^{W,u,k}$ defined in (\ref{eqn:wdmcap})
\begin{eqnarray}  \label{DWuk:eqn}
 D^{W,u,k,E} &\approx &
   \tau_T + 2 \tau_T
      +\frac{\rho^{T,d,k}\bar{L}}{2C^k_T}
      +\Phi\left(\frac{\lambda^{W,u,k}}{W^k C_{W}^k}\right)
         + \tau_T + \frac{\bar{L}}{C_W}.
\end{eqnarray}

\subsubsection{Empty Carrier for Upstream Data Transmission}
With switching between upstream and downstream transmission, the
combined upstream and downstream traffic has to be accommodated
on the $C^W_k$ WDM channels, resulting in the load
\begin{eqnarray}
\rho^{W,k}=\frac{ \lambda^{W,d,k} + \lambda^{W,u,k}}
           {W^k C_W^k}.
\end{eqnarray}
which is served out of a (virtual) queue holding both upstream and
downstream traffic. The resulting queueing delay is approximately
$\Phi( \rho^{W,k})$.

If an empty carrier is used for upstream data transmissions, a
waiting period equal to the one-way propagation delay $\tau_T$ is
introduced when switching a WDM channel from upstream to downstream
transmission, or vice versa. More specifically, when switching from
downstream to upstream transmission, once the downstream
transmission has ended, the immediately subsequently transmitted
carrier signal takes $\tau_T$ to reach the ONU. Once the carrier
signal starts to arrive at the ONU, it can immediately commence its
upstream data transmission. Similarly, when switching from upstream
to downstream transmission, the last bit of the upstream data
transmission requires $\tau_T$ to reach the OLT. Only when the last
bit of the upstream transmission has reached the OLT, can the OLT
commence a new downstream data transmission.

We denote
by $p_{s,k}$ the probability that a `switchover' between
upstream and downstream transmission, or vice versa, takes place
before a data transmission on the WDM channels at CO $k$.
Consider the superposition of two
independent sequences of Poisson arrival times with rates $\lambda_1$
and $\lambda_2$, respectively. Let $P_1$ be the event of
an arrival from the first process (resp.\ denote $P_2$ for an arrival
of the second process) and let $SO$ denote the event
that a switchover occurs. Then, the probability for a switchover equals
\begin{eqnarray}
\pr{SO}=\pr{SO|P_1} \pr{P_1} + \pr{SO|P_2} \pr{P_2} =
\frac{\lambda_2}{\lambda_1+\lambda_2} \cdot
\frac{\lambda_1}{\lambda_1+\lambda_2} +
\frac{\lambda_1}{\lambda_1+\lambda_2} \cdot
\frac{\lambda_2}{\lambda_1+\lambda_2}.
\end{eqnarray}
Note that the probability that a switchover
occurs, given that we consider an arrival of the first process, is
equal to the probability that an exponential random variable with mean
$1/\lambda_2$ is smaller than another (independent) exponential random
variable with mean $1/\lambda_1$.
Simplifying, we obtain
\begin{equation}
p_{s,k}=\frac{2 \lambda_1 \lambda_2}{(\lambda_1+\lambda_2)^2}, \label{eqn:noofswitchovers}
\end{equation}
whereby, noting that we consider CO $k$, we have the downstream traffic
rate $\lambda^{W,d,k}$ and the
upstream traffic rate $\lambda^{W,u,k}$ sharing the
$W^k$ channels, each with capacity $C_W^k$, i.e.,
$\lambda_1:= \lambda^{W,d,k} / (W^k C_W^k)$
and $\lambda_2:= \lambda^{W,u,k} / (W^k C_W^k)$.

For each switchover we calculate a loss of
transmission time of $\tau_T$.
We model the
switchovers as changes in the packet length distribution:
the packet transmission time is extended by
$C_W^k \tau_T$ with probability $p_{s,k}$.
This gives a new mean of the packet
distribution of $\bar{L}+ p_{s,k} C_W^k \tau_T$ and second moment
$\sigma_L^2 + 2 \bar{L} p_{s,k} C_W^k \tau_T + p_{s,k}
( C_W^k \tau_T)^2$.
Using this modified message length distribution, we
obtain the delay on the WDM channels as
\begin{eqnarray}
D^{W,d,k} &=&
   \Phi( \rho^{W,k} ) + \tau_T + \frac{\bar{L}}{C_T^k} - B^{W,d,k}.
  \label{e:downquingwdm}
\end{eqnarray}

The upstream traffic experiences additional delay components:
\begin{itemize}
\item The residual time of the upstream TDM channel
$\tau_T$ to account for the delay
from packet generation
until transmission of the corresponding report.
\item the round trip propagation delay $2 \tau_T$ to account for the
upstream propagation of the report and downstream propagation of the
grant.
\item the queueing delay for the grant prior to its
transmission on the downstream TDM channel, which is
$\Phi( \rho^{T,d,k})$ without priority for grants,
and $\rho^{T,d,k}\bar{L}/2$ with priority for the grant messages
(considered in (\ref{DWuk_empty:eqn}).
\end{itemize}
\begin{equation} \label{DWuk_empty:eqn}
    D^{W,u,k}:=\tau_T
       + \frac{\rho^{T,d,k}\bar{L}}{2C_T^k} + 2 \tau_T + D^{W,d,k}.
\end{equation}

\subsection{Delay on GPON }

Let $\delta$ denote the frame duration of 125 $\mu$s of the GPON.
Consider a packet being generated at an ONU attached to OLT $k$. The
packet has to wait on average $\delta/2$ for the beginning of the
next frame in which it will be included in a dynamic bandwidth
report (DBRu) field. This next frame has a duration (transmission
delay) of $\delta$ and takes $\tau_T$ to propagate to the OLT.

Once
arrived at the OLT, the bandwidth report has to be processed by the
OLT and the grant to the ONU for the packet's transmission is
included in the bandwidth map (BWmap) of the next downstream frame.
Even with negligible processing time at the OLT, there is a delay of
up to $\delta$ until the beginning of the next downstream frame.
More specifically,
let $\omega,\ 0 \leq \omega < \delta$, [in seconds]
denote the offset between the up and down channels defined as
follows. At the instant when a new slot starts on the upstream
channel, $\omega$ seconds have passed of the current
downstream channel slot, i.e., for $\omega = 0$, the slots on the
upstream and downstream channels are aligned.
Then, the time until the beginning of the downstream frame containing
the BWmap of the considered packet is
\begin{eqnarray}
\gamma_1 =  \left( \left \lfloor \frac{\tau_T - \omega}{\delta} \right \rfloor 
  + 1 \right) 
   \delta - (\tau_T - \omega) \\
 = \left ( 1 - \left ( \frac{\tau-\omega}{\delta} - \left \lfloor
\frac{\tau-\omega}{\delta} \right \rfloor \right ) \right ) \delta.
\end{eqnarray}

The downstream frame has a transmission delay of $\delta$ and
propagation delay of $\tau_T$. The packet has to wait for 
\begin{eqnarray}
\gamma_2 = \left( \left \lfloor \frac{\tau_T + \omega}{\delta} \right \rfloor 
     + 1 \right) 
   \delta - (\tau_T + \omega) \\
= \left ( 1 - \left ( \frac{\tau+\omega}{\delta} - \left \lfloor
\frac{\tau+\omega}{\delta} \right \rfloor \right ) \right ) \delta 
\end{eqnarray}
until the beginning of
the next upstream frame before it can possibly be transmitted. Thus,
it takes overall on average $\delta/2 + \tau_T + \gamma_1 + \tau_T +
\gamma_2$ from the instant the packet is generated to the instant
the packet becomes eligible for upstream transmission. And then, the
packet is put into a general queue for the upstream channel. In
terms of the mean packet delay, this channel can be modeled as an
M/G/1 queue (noting that the specific scheduling discipline does not
affect the overall mean packet delay in the GPON, as long as the
channel is operated in work conserving manner, i.e., is not left
idle while packets are queued), with corresponding delay
$\Phi(\rho^{T,u,k})$. Finally, the packet experiences the
transmission delay $\bar{L} / C_T^k$ and propagation delay $\tau_T$.
Overall, the mean delay for the TDM upstream channel is
\begin{eqnarray}
D^{T,u,k,G} = \frac{5 \delta}{2} + \gamma_1 + \gamma_2 +
\Phi(\rho^{T,u,k}) + 3 \tau_T + \frac{\bar{L}}{C_T^k}.
\end{eqnarray}
Analogously, we obtain for the WDM upstream channels which experience
the same delays for the report-grant cycle but carry the load
$\lambda^{W,u,k} / (W^k C^k_W)$:
\begin{eqnarray}
D^{W,u,k,G} = \frac{5 \delta}{2} + \gamma_1 + \gamma_2 +
\Phi(\frac{\lambda^{W,u,k}}{W^k C^k_W}) + 3 \tau_T + \frac{\bar{L}}{C_T^k}.
\end{eqnarray}
Note that $\omega$ can be adjusted to save up to one $\delta$ of delay.
Specifically, there are two cases in the
minimization of $\gamma_1 + \gamma_2$:

(A) If
\begin{eqnarray}
\tau/\delta - \lfloor \tau/\delta \rfloor < \lfloor \tau/\delta \rfloor+1
- \tau/\delta,
\end{eqnarray}
then any $\omega$ with
\begin{eqnarray}
\tau/\delta - \lfloor \tau/\delta \rfloor < \omega/\delta < \lfloor
\tau/\delta \rfloor+1 - \tau/\delta
\end{eqnarray}
is optimal.

(B) If
\begin{eqnarray}
\tau/\delta - \lfloor \tau/\delta \rfloor > \lfloor \tau/\delta \rfloor+1
- \tau/\delta,
\end{eqnarray}
then any $\omega$ with
\begin{eqnarray}
0\leq \omega/\delta < \lfloor \tau/\delta \rfloor+1 - \tau/\delta
\end{eqnarray}
is optimal.

Turning to the downstream transmission, we note that a packet
arriving at the OLT has to wait on average $\delta/2$ for the beginning
of the next downstream frame, i.e., before it becomes eligible for transmission.
The packet also experiences the average downstream queueing delay
$\Phi(\rho^{T,d,k})$, transmission delay, propagation delay, and
delay correction analogous to (\ref{Dtdk:eqn})
for a total delay of approximately
\begin{equation} \label{DTdkG:eqn}
D^{T,d,k,G} := \frac{\delta}{2} +
  \Phi \left( \rho^{T,d,k} \right) +
    \tau_T + \frac{\bar{L}}{C_T^k} - B^{T,d,k},
\end{equation}

The delay for the downstream WDM channels is obtained by replacing
$\Phi \left( \rho^{T,d,k} \right)$
by $\Phi \left( \frac{\lambda^{W,d,k}}{W^k C^k_W} \right)$
in (\ref{DTdkG:eqn}).

\subsection{Delay in the Ring/PSC Star Subnetwork}
We first evaluate the packet delay in the PSC star subnetwork as follows.
With $\tau_P^f$ denoting the frame duration [in seconds]
on the PSC, a newly arrived packet at the CO waits on average
$\tau_P^f/2$ before its control packet can be sent.
The control packet experiences a propagation delay of
$\tau_{P}$.
Once the control packet is received, the packet enters the virtual
queue for the destination CO. This queue experiences a load of
$\rho^{P,l} = \lambda^P(l) / C_{P}$ with $\lambda^P(l)$ given in
(\ref{lambdaPl:eqn}).
Adding in the transmission and propagation delays of the data packet
over the PSC, we obtain
\begin{equation}
 D^{P}(l) = \frac{1}{2}\,\tau_{P}^{f} + \tau_{P}
+ \Phi(\rho^{P,l}) + \tau_{P} + \frac{\bar{L}}{C_{P}} - B^{P,l},
\end{equation}
where $B^{P,l}$ is a correction term given by
\begin{equation}
 B^{P,l} = \sum_{\text{CO $k$}} \left[\sum_{\text{RPR $m$ adjacent to CO $k$}} \Phi(\rho^{RPR,P,m,k,l}) + \Phi(\rho^{T,P,k,l}) + \Phi(\rho^{W,P,k,l})\right]
\end{equation}
with
\begin{equation}
 \rho^{m,R,P,k,l} = \sum_{m,j\in \mathcal{N}} p_{mj}(m,k) p_{mj}(k,l) T(m,j)
    \bar{L} / C_{RPR}
\end{equation}
denoting the traffic that
originates at RPR node $m$ and flows over the PSC from CO $k$ to CO $l$.
Analogously, we define
\begin{equation}
 \rho^{T,P,k,l} = \sum_{i \in \mathcal{C}_T^k,j\in \mathcal{N}} p_{ij}(k,l) T(i,j) \bar{L} / C_{T}
\end{equation}
and the respective quantity $\rho^{W,P,k,l}$, where $\mathcal{C}_T^k$ is replaced by $\mathcal{C}_W^k\cup\mathcal{C}_L^k$ and $C_{T}$ by $C_{W}^k$.

The packet delay in
the ring $D^R_{ij}$ from a CO/hotspot $i$
to another CO/hotspot (or a destination ring node) $j$
is given by Eqn.\ (22) in \cite{MHSR05} with
last two sums replaced by
\begin{equation}
   D^{P}_{ij} = \sum_{k,l} p_{ij}(k,l)
      D^{P}(l).
\end{equation}
Note that all links $(k,l)$ that are not used for the
transmission from node $i$ to node $j$ have $p_{ij}(k,l)=0$ in the sum above.
The packet delay from an RPR ring node to a CO/hotspot
(or a destination RPR ring node)
is given by Eqn.\ (21) in \cite{MHSR05} with the last sum
replaced by $D^{P}_{ij}$.

\subsection{Delay in AWG Star Subnetwork}
The packet delay in the AWG star subnetwork consists of the following
components:
\begin{itemize}
\item The residual cycle time of the upstream TDM channel
$\tau_T$ to account for the delay
from packet generation
until transmission of the corresponding report.
\item The round trip propagation delay $2 \tau_T$ to account for the
upstream propagation of the report and downstream propagation of the
grant.
\item The queueing delay for the grant prior to its
transmission on the downstream TDM channel, which is
$\Phi( \rho^{T,d,k})$ without priority for grants,
and $\rho^{T,d,k}\bar{L}/2$ with priority for the grant messages
(considered in (\ref{DAkl:eqn})).
\item The queueing delay due to several LR ONUs with total
load $\rho^A (k,l) = \lambda^A(k,l) /(c(k,l) C_{A})$ sharing
the $c(k,l)$ channels from CO $k$ to CO $l$.
\item The average packet transmission delay $\bar{L} / C_A$
and propagation delay $\tau_A$.
\end{itemize}
Thus, approximately,
\begin{equation} \label{DAkl:eqn}
D^A(k,l) = 
   + 3\tau_T + \frac{\rho^{T,d,k}\bar{L}}{2C_T^k}
  + \Phi(\rho^A (k,l)) +\tau_{A} + \frac{\bar{L}}{ C_A }.
\end{equation}
Note that the first two terms in the this expression only occur
because the LR ONUs have to use the upstream TDM channel to
register their packets.
By averaging over all channels we obtain the average packet delay
on the AWG star subnetwork:
\begin{equation}
    D^A = \sum_{\text{CO $k$, $l$}}
        D^A(k,l) \cdot \frac{ \lambda^A(k,l) }{\sum_{\text{CO $k'$, $l'$}}
            \lambda^A(k',l')}.
\end{equation}

\subsection{The Hotspots}
Note that the hotspot nodes do not require any special analysis. The
hotspots simply have typically a higher traffic volume, which is
expressed in the respective $T(i,j)$ and $T^A(i,j)$. A hotspot node
is modeled as any other RPR ring node (or CO) in the ring/PSC star
subnetwork, and as a CO in the AWG star subnetwork.

\subsection{Overall Delay}
We obtain the overall average packet delay by weighing
the different paths according to their packet traffic rates.
First, for traffic transmitted over the ring/PSC star subnetwork:
\begin{equation}
D^{R,P}=\sum_{i,j} D(i,j) \, \frac{T(i,j)}{\sum_{i',j'} T(i',j')},
\end{equation}
where
\begin{itemize}
\item $D(i,j)=D^R_{ij}$ for traffic from RPR ring node/hotspot $i$
     to RPR ring node/hotspot $j$.
\item $D(i,j)=D^R_{il} + D^{T,d,l}$ for traffic from
        RPR ring node/hotspot $i$
    to TDM ONU $j$ at CO $l$ (resp.\ $D^{W,d,l}$
     for traffic to WDM or LR ONU at CO $l$).
\item $D(i,j)=D^{T,u,k} + D^R_{kl} + D^{T,d,l}$
    for traffic from TDM ONU $i$ at CO $k$ to TDM ONU $j$ at CO $l$
     (resp.\ $D^{W,u,k}$ for traffic from WDM or LR ONU $i$
   at CO $k$ and resp.\ $D^{W,d,l}$ for traffic to
   WDM or LR ONU $j$ at CO $l$).
 Note that for intra-CO traffic from an ONU $i$ to another ONU $j$
 attached to the same CO $k$, $D(i,j)$ gives the intra-CO as
   $D^{T,u,k} + 0 +D^{T,d,k}$ (since the ring delay is zero for $k=l$);
  the scenarios with intra-CO traffic to and/or from a WDM ONU are
  captured analogously.
\item $D(i,j)=D^{T,u,k}+ D^R_{kj}$ for traffic from TDM ONU $i$ at CO $k$ to
    RPR ring node/hotspot $j$
    (resp.\ $D^{W,u,k}$ for traffic from WDM or LR ONU $i$ at CO $k$).
\end{itemize}
Overall we obtain:
\begin{eqnarray} D = D^{R,P} \,\frac{\sum_{i,j} T(i,j)}
      {\sum_{i,j} T(i,j) + \sum_{\text{LR ONUs $i$,$j$}}T^A(i,j)}
        + D^A \, \frac{\sum_{\text{LR ONUs $i$,$j$}}
        T^A(i,j)}{\sum_{i,j} T(i,j)
    + \sum_{\text{LR ONUs $i$,$j$}} T^A(i,j)}.
\end{eqnarray}

\section{Numerical and Simulation Results}
\label{num:sec}

This section presents numerical results on the throughput-delay
performance, first for isolated PONs and then for integrated
access-metro networks,
obtained from our analysis and extensive verifying simulations with
95\% confidence intervals. The propagation speed is set to
$2\cdot 10^8$ m/s. We first consider uniform traffic where each ONU
generates the same amount of packet traffic with a packet size
randomly uniformly distributed over $[64,1518]$ bytes.
In the context of an isolated PON, a packet
generated by a given ONU is destined to any of the other $N-1$ ONUs of
the same PON with equal probability $1/(N-1)$.

\subsection{Isolated PONs}
\label{isopon:sec}
\begin{figure}[t]
\begin{center}
\includegraphics[width=.5\textwidth]{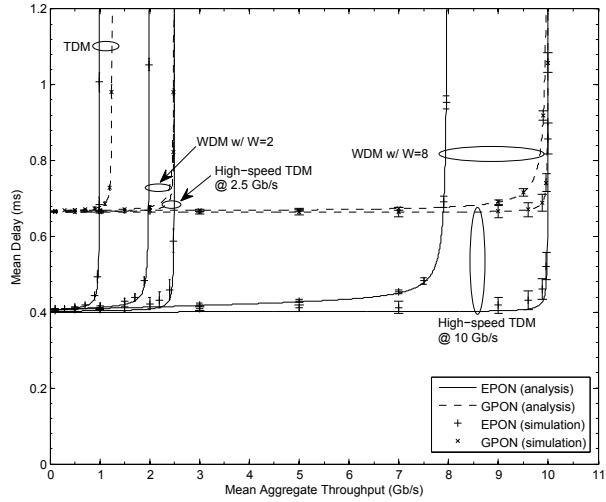}
\caption{Mean delay $D$ on upstream TDM/WDM channels
of high-speed TDM and WDM EPON/GPON vs.\ mean aggregate throughput $r_T$ 
for $N_T=32$ TDM ONUs and $N_W=32$ WDM ONUs, respectively.}
\label{fig:res2}
\end{center}
\end{figure}

Fig.~\ref{fig:res2} compares the mean delay $D$ on the upstream TDM/WDM
channels of conventional TDM, high-speed TDM, and WDM EPON/GPON
networks vs. the mean aggregate throughput $r_T$. 
We consider an EPON with $C_T=1$ Gb/s and a (symmetric) GPON with $C_T=1.25$
whereby the ONUs
are located at 20 km from the OLT.
We consider a fixed number of
$N_T=32$ TDM ONUs and $N_W=32$ WDM ONUs, respectively. Both high-speed
TDM PONs operate at a data rate of $C_T\in\{2.5, 10\}$ Gb/s. In
addition to the pair of legacy TDM wavelength channels, the WDM EPON
and WDM GPON deploy $W\in\{2,8\}$ wavelength channels via
remodulation, each operating at $C_W=C_T=1$ Gb/s and $C_W=C_T=1.25$
Gb/s, respectively.
We observe that the EPON achieves significantly lower delays
than the GPON at small to medium
traffic loads. This EPON advantage is due to its
underlying variable-length polling cycle
compared to the fixed length framing structure of the GPON.
We further observe that analysis and simulation results match very
well, except that the analysis underestimates
the mean EPON delay slightly at medium traffic loads.

\begin{figure}[t]
\begin{center}
\includegraphics[width=.5\textwidth]{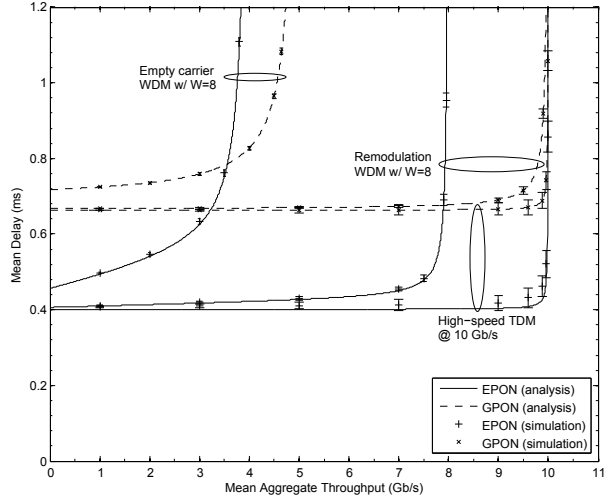}
\caption{Mean delay $D$ vs. mean aggregate throughput $r_T$ 
of WDM EPON/GPON using remodulation and empty carrier.}
\label{fig:res3}
\end{center}
\end{figure}
Fig.~\ref{fig:res3} shows the 10 Gb/s high-speed TDM and WDM EPON/GPON
with remodulation of Fig.~\ref{fig:res2} and compares them to a WDM
EPON/GPON using the commercially available empty carrier
approach~\cite{Lin08}. The empty carrier approach severely
deteriorates the performance of both WDM EPON and WDM GPON, suffering
from a higher mean delay and a significantly lower mean aggregate
throughput on the upstream TDM/WDM channels than a WDM EPON/GPON based
on remodulation. This is due to the fact that in the empty carrier
approach each WDM wavelength channel is used for bidirectional
transmission, i.e., upstream and downstream transmissions alternate,
as opposed to remodulation where upstream transmissions
are not delayed by downstream transmissions.

\begin{figure}[t]
\begin{center}
\includegraphics[width=.5\textwidth]{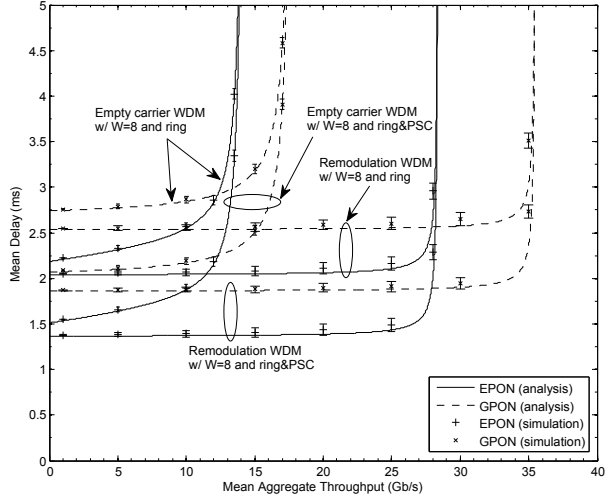}
\caption{Mean delay $D$ vs. mean aggregate throughput $r_T$
of three WDM PONs interconnected through $(i)$ a ring,
or $(ii)$ a ring combined with a $4\times 4$ PSC.}
\label{fig:res4}
\end{center}
\end{figure}

\subsection{Integrated Access-Metro Networks}
\label{intaccmet:sec}
Next, we investigate different methods to interconnect multiple
high-speed TDM/WDM PONs by means of a ring, PSC, and/or
AWG. Fig.~\ref{fig:res4} depicts the mean delay vs. mean aggregate
throughput of three of the aforementioned WDM EPONs/GPONs
interconnected through either $(i)$ a ring only, or $(ii)$ a ring in
conjunction with a $4\times 4$ PSC (i.e., $P=4$),
for shortest path (minimum hop) routing. The circumference of the
bidirectional ring is set to 100 km and it comprises $N_r=4$ equally
spaced ring nodes. Both ring and PSC operate at a data rate of 10
Gb/s, i.e., $C_R=C_P=10$ Gb/s. We consider
uniform source and
destination traffic originating
from and going to any of the $3\cdot 32=96$ WDM ONUs, $N_r = 4$ ring nodes,
and $H=1$ remote CO (see Fig.~\ref{fig:AWG}).
Formally, for uniform source traffic,
$\sigma(i) = \sigma\ \forall i \in \mathcal{N}$ and the total
traffic bit rate in the network is
$r_T = \eta \bar{L} \sigma$.
As shown in
Fig.~\ref{fig:res4}, using a PSC that provides short-cut links to the
ring helps decrease the mean delay considerably, but in the
considered example network configuration does not
lead to an increased mean aggregate throughput. (Similar
observations were made for 10 Gb/s high-speed TDM PONs, not shown here
due to space constraints.)
In particular, for the EPON with
uniform source and uniform destination traffic,
both the upstream (Eqn.~(\ref{eqn:wdmcap})) and the downstream WDM
channel with remodulation capacity constraint
(Eqn.~(\ref{lambda_WDM_down:eqn}))
give the bound
\begin{eqnarray} \label{rt_WDMup:eqn}
r_T < \frac{\eta (\eta - 1) (W + 1) C}
    {(\eta-1) (N_T + N_W) + \eta_{TWr} N_L}
\end{eqnarray}
with $\eta_{TWr} = (P-H) (N_T + N_W) + N_r$ denoting the total
number of TDM/WDM ONUs and ring nodes.
For the considered scenarios with $N_W = 32$ WDM ONUs (and no
TDM or LR ONUs) this bound
reduces to
$r_T < \eta (W + 1) C / N_W = 28.4$ Gbps, which is
lower than the bounds imposed by the ring and PSC, as
detailed next,
and hence governs the maximum mean aggregate throughput.

For further analysis of the ring/PSC stability condition
(\ref{lambdaPl:eqn}), we fix the ring network to the structure
illustrated in Fig.~\ref{fig:AWG} with $N_r = 4$ ring nodes, 
$P = 4$ OLTs (of which $H = 1$ is a remote CO), and with one ring node
between two OLTs. The highest traffic rate on a PSC channel arises
for the uniform traffic pattern on the channel toward an OLT 
when the remote CO is two ring hops from the
considered OLT. We refer henceforth to the considered OLT as the
``target OLT'', the OLT opposite of the target OLT around the ring as
the ``opposite OLT'', and the OLT situated two hops along the ring
from the target OLT as the ``adjacent OLT'' (which is located opposite
the remote CO). We have the following contributions to the traffic
load on the PSC home channel of the target OLT:
\begin{itemize}
\item The opposite OLT has
$N_T + N_W$ TDM/WDM ONUs sending
$(i)$ to the $N$ nodes in the EPON attached to the target OLT, plus
$(ii)$ with probability one half to the
two ring nodes situated one hop
from the target OLT over the target home channel.
In addition, the opposite OLT has $N_L$ LR ONUs
sending
$(i)$ to the $N_T + N_W$ TDM/WDM ONUs in the EPON attached
to the target OLT, plus
$(ii)$ with probability one half to the
two ring nodes situated one hop
from the target OLT over the target channel.
Thus, the opposite OLT contributes
\begin{eqnarray}
\frac{r_T}{\eta} (N_T + N_W) \frac{N + \frac{1}{2} 2}{\eta-1} +
   \frac{r_T}{\eta} N_L \frac{N_T + N_W + \frac{1}{2} 2}{\eta-1}.
\end{eqnarray}
\item The adjacent OLT has the same contribution as the opposite OLT,
except that only the traffic to EPON nodes attached to the 
target OLT is sent over the PSC; the ring nodes situated
one hop from the target OLT are reached with one hop over the ring
(versus two hops over the PSC and then ring).
Furthermore, the remote CO contributes as much as one
LR ONU at the adjacent CO.
Overall, the contribution from the adjacent OLT and the remote CO is thus
\begin{eqnarray}
\frac{r_T}{\eta} (N_T + N_W) \frac{N }{\eta-1} +
   \frac{r_T}{\eta} (N_L+1) \frac{N_T + N_W}{\eta-1}.
\end{eqnarray}
\item The two ring nodes situated within one ring hop from the
target OLT do not send traffic toward the target OLT over the PSC.
\item The two ring nodes situated three hops in either ring direction
from the target OLT send to all ONUs attached to the target OLT over
the PSC (and directly over the ring to the ring nodes one hop
from the target OLT) contributing
\begin{eqnarray}
\frac{r_T}{\eta} 2 \frac{N}{\eta - 1}.
\end{eqnarray}
\end{itemize}
Combining these contributions results in the constraint
\begin{eqnarray} \label{rt_PSC:eqn}
r_T &<& \frac{\eta (\eta - 1) C_P}{d_1} \\
 d_1 &=& (N_T + N_W ) ( 3N + 1 ) + N_L (3 N_T +  3N_W + 1) \nonumber \\
    &&\ \ \ \ \ + 2 N + N_T + N_W .
\end{eqnarray}
For our scenario with $N_L = 0$, $N_T = 0$ and $N = N_W$ 
this simplifies to
\begin{eqnarray} \label{rt_PSCscen:eqn}
r_T < \frac{\eta (\eta - 1) C_P}
  { 3 N_W  ( N_W + 4/3 )} = 31.5625 \mbox{Gbps}.
\end{eqnarray}

\begin{figure}[t]
\begin{center}
\includegraphics[width=.5\textwidth]{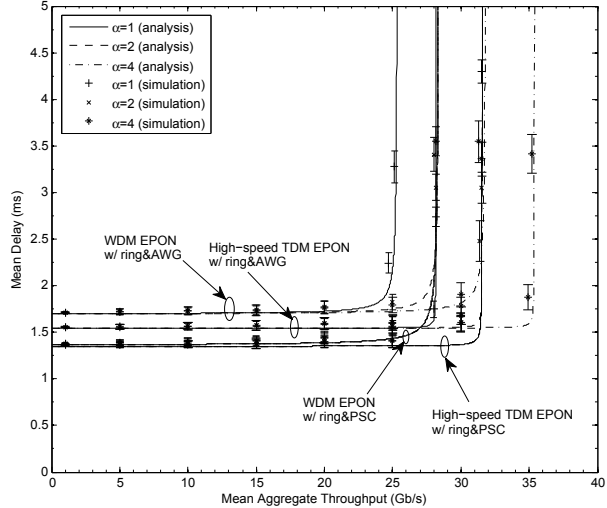}
\caption{Mean delay $D$ vs. mean aggregate throughput $r_T$
of three WDM EPONs and three high-speed TDM EPONs interconnected
through $(i)$ a ring\&PSC, or $(ii)$ a ring\&AWG for different
source traffic non-uniformity $\alpha\in\{1,2,4\}$.}
\label{fig:res5}
\end{center}
\end{figure}
In the following, we study the impact of non-uniform traffic on the
throughput-delay performance of NG-PONs. Let us first focus on
\textit{non-uniform source traffic}, where nodes generate different
traffic rates. For now, we continue to consider uniform destination
traffic. More specifically, $N_m$ of the ONUs in each NG-PON as well
as all $N_r$ ring nodes generate traffic at a medium bit
rate of $\sigma \bar{L}$. Furthermore, we introduce a source traffic
non-uniformity $\alpha$, $\alpha\geq 1$, and let $N_l$ lightly loaded
ONUs in each NG-PON generate traffic at a low bit rate of
$\sigma\bar{L}/\alpha$, and $N_h$ highly loaded ONUs in each NG-PON as
well as the remote CO generate traffic at a high bit rate of
$\alpha\sigma\bar{L}$. Note that $\alpha=1$ denotes uniform traffic,
which has been studied above.

Fig.~\ref{fig:res5} compares the mean
delay vs. mean aggregate throughput performance of three WDM EPONs
with $W=8$ wavelengths in remodulation mode, each operating at 1 Gb/s,
with that of three 10 Gb/s high-speed TDM EPONs, interconnected with
the remote CO through a ring in conjunction with $(i)$ a $4\times 4$
PSC, or $(ii)$ a $4\times 4$ AWG using $\Lambda_{\rm AWG}=4$
wavelengths. The ring, PSC, and AWG operate at 10 Gb/s,
i.e., $C_R=C_P=C_A=10$ Gb/s. In each EPON, we set $N_l=16$ and
$N_m=N_h=8$ and consider different source traffic non-uniformity
$\alpha\in\{1,2,4\}$.
In the ring\&PSC configuration, there are $N_W = 32$ WDM ONUs in a
given WDM EPON (resp.\ $N_T= 32$ ONUs in a high-speed TDM EPON).
In the ring\&AWG configuration, the $N_h$ highly
loaded ONUs are upgraded to LR-ONUs in each high-speed WDM EPON
(and each high-speed TDM EPON which is connected with $P$ high-speed
wavelength channels to the AWG).

We observe from Fig.~\ref{fig:res5} that
the ring\&PSC configurations are insensitive to source traffic
non-uniformities.
This is because the shift in traffic generation from lightly to
heavily loaded ONUs with increasing source traffic non-uniformity
$\alpha$ does not significantly shift the portion of the
total network traffic load that needs to traverse the EPON downstream
WDM channels.
In contrast, for the ring\&AWG configurations we observe from
Fig.~\ref{fig:res5}
increases in the aggregate network throughput as the traffic becomes more
non-uniform.
With increasing $\alpha$, the heavily loaded ONUs account for
a larger portion of the total network traffic.
Thus, the traffic portion that can be off-loaded from the
EPON WDM channels to the AWG channels, namely all traffic
between pairs of heavily loaded ONUs, increases with $\alpha$,
resulting in an increased aggregate throughput.
(Similar
observations were made for high-speed TDM and WDM GPONs.)

The total traffic bit rate in the network is
\begin{eqnarray}
r_T = \left [ (P - H) \left(\frac{N_l}{\alpha} + N_m + \alpha N_h \right)
       + N_r + \alpha H \right] \sigma \bar{L}.
\end{eqnarray}
For notational convenience, we define
the equivalent number of medium bit rate traffic nodes as
\begin{eqnarray}
\eta_{\alpha} = (P - H) \left(\frac{N_l}{\alpha} + N_m + \alpha N_h \right)
       + N_r + \alpha H,
\end{eqnarray}
with which we express the medium traffic bit rate in terms of
$r_T$ as $\sigma \bar{L} = r_T / \eta_{\alpha}$.
We focus first on the WDM EPON with ring\&PSC scenario.
Note that $\lambda^{T,u,k} = \lambda^{T,d,k} = 0$.
For the upstream WDM channels we obtain
by noting that all traffic generated at the nodes of a given EPON has
to go up on the WDM (and TDM) channels
\begin{eqnarray}
\lambda^{W,u,k} = \frac{r_T}{\eta_{\alpha}}
 \left( \frac{N_l}{\alpha} + N_m + \alpha N_h \right)
\end{eqnarray}
and the corresponding limit
\begin{eqnarray}  \label{rt_WDMupaup:eqn}
r_T < \frac{\eta_{\alpha} (W + 1) C}
       { \frac{N_l}{\alpha} + N_m + \alpha N_h}.
\end{eqnarray}
For the considered scenario with $\eta = 101$,
$\eta_1 = 101$, $\eta_2 = 102$, and $\eta_4 = 140$ we obtain
the constraints
$r_T < 28.40625$ for $\alpha = 1$,
$r_T < 28.6875$ for $\alpha = 2$, and
$r_T < 28.636364$ for $\alpha = 4$.

For the downstream WDM (and TDM) channels using remodulation
we note that all traffic
generated by one of the $N$ nodes at the considered EPON and destined to any
of the other $N-1$ nodes at the EPON as well as any traffic generated
by one of the other nodes in the network and destined to any of the
$N$ nodes at the EPON contributes to the downstream load
\begin{eqnarray}
\lambda^{W,d,k} = \frac{r_T}{\eta_{\alpha}}
 \left[ \frac{N_l}{\alpha}  \frac{N - 1}{\eta-1} + N_m \frac{N-1}{\eta-1}
       + \alpha N_h \frac{N-1}{\eta-1} \right. \nonumber \\
      \left. + (P-H-1) \frac{N_l}{\alpha}  \frac{N - 1}{\eta-1}
       + (P-H-1) N_m  \frac{N }{\eta-1} \right. \nonumber \\
       \left. + (P-H-1) \alpha N_h \frac{N}{\eta-1} + N_r \frac{N }{\eta-1}
      + \alpha H \frac{N }{\eta-1}   \right],
\end{eqnarray}
resulting in the limit
\begin{eqnarray}  \label{rt_WDMdownaup:eqn}
r_T < \frac{\eta_{\alpha} (\eta-1) (W+1)C}
   { d }.
\end{eqnarray}
with
\begin{eqnarray}
d = (N-1) \left( \frac{N_l}{\alpha} + N_m +  \alpha N_h \right) \nonumber \\
    + N \left( (P-H-1) \left( \frac{N_l}{\alpha} + N_m +  \alpha N_h \right)
   \right. \nonumber \\
        \left. + N_r + \alpha H \right)
\end{eqnarray}
For the considered scenario we obtain the limits
$r_T < 28.40625$ for $\alpha = 1$,
$r_T < 28.40346$ for $\alpha = 2$, and
$r_T < 28.40397$ for $\alpha = 4$.
Taken together, these capacity results confirm the simulation
results for the ring\&PSC configuration.

Next, we focus on the WDM EPON with ring\&AWG.
For the upstream WDM channels note that
the $N_l$ low traffic ONUs and the $N_m$ medium traffic ONUs at a given
EPON send all generated traffic upstream on the WDM
(and TDM) channel. In addition the $N_h$ high traffic ONUs send the
traffic to nodes not connected to the AWG upstream on the WDM channels, i.e.,
\begin{eqnarray}
\lambda^{W,u,k} = \frac{r_T}{\eta_{\alpha}}
   \left( \frac{N_l}{\alpha} + N_m +
        \alpha N_h \frac{\eta_{TWr}}{\eta-1} \right)
\end{eqnarray}
with \begin{eqnarray}
\eta_{TWr} = (P-H) \left( N_l + N_m \right) + N_r
\end{eqnarray}
denoting the number of nodes not connected to the AWG.
The corresponding capacity limit is
\begin{eqnarray}  \label{rt_WDMupaup:eqn}
r_T < \frac{\eta_{\alpha} (W + 1) C}
       {\left( \frac{N_l}{\alpha} + N_m +
      \alpha N_h \frac{\eta_{TWr}}{\eta-1} \right) }.
\end{eqnarray}
We obtain for our scenario for which $\eta_{TWr} = 76$, for
$\alpha = 1$ $r_T < 30.2194$,
for $\alpha = 2$ $r_T < 32.5994$, and for
$\alpha = 4$ $r_T < 34.6916$.

For the AWG channels we obtain the highest load
on the channels connecting two EPONs, namely
\begin{eqnarray}
\lambda^A = \frac{r_T}{\eta_{\alpha}} \alpha N_h \frac{N_h}{\eta-1}
\end{eqnarray}
and the corresponding limit
\begin{eqnarray}   \label{rt_AWGaup:eqn}
r_T < \frac{\eta_{\alpha} (\eta-1) c C}
        {\alpha N_h^2}.
\end{eqnarray}
For our scenario, $r_T < 1578.125$ for $\alpha = 1$, $r_T < 796.875$
for $\alpha = 2$, and $r_T < 453.125$ for $\alpha = 4$.

\begin{figure}[t]
\begin{center}
\includegraphics[width=.5\textwidth]{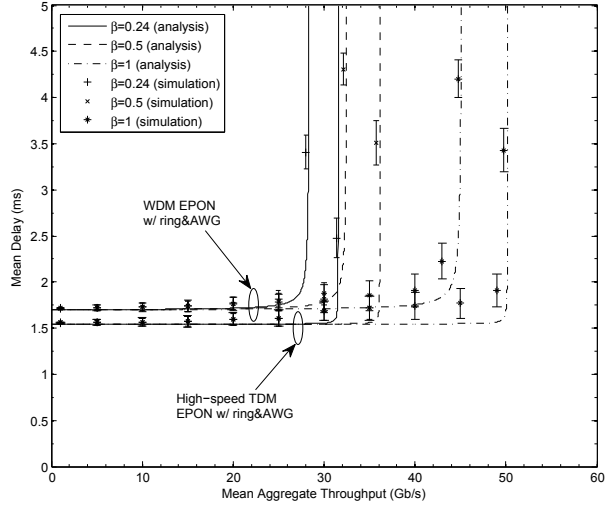}
\caption{Mean delay $D$ vs. mean aggregate throughput $r_T$ of three
WDM EPONs and  three high-speed TDM EPONs interconnected through
a ring\&AWG for $\alpha=2$ and different destination traffic
non-uniformity $\beta\in\{0.24,0.5,1\}$.}
\label{fig:res6}
\end{center}
\end{figure}
Fig.~\ref{fig:res6} considers the ring\&AWG configurations of the
previous figure for a fixed source traffic non-uniformity $\alpha=2$
and illustrates the impact of
\emph{non-uniform destination traffic}.
Specifically, a packet generated by an LR ONU or CO is
destined to another LR ONU/CO with probability
$(\eta_{LH}-1)/(\eta-1)\leq\beta\leq 1$, whereby
$\eta_{LH}=(P-1)N_h+H$ denotes the number of LR-ONUs and CO
in the network.
Note that in our case $\beta=(\eta_{LH}-1)/(\eta-1)=0.24$
corresponds to uniform destination traffic.
We observe from Fig.~\ref{fig:res6} that both AWG network configurations
are quite sensitive to
destination traffic non-uniformity. The maximum mean aggregate
throughput is significantly increased
as the fraction of traffic routed over the AWG increases.

We illustrate the application of the
capacity constraints of Sections~\ref{treecap:sec}--\ref{awgcap:sec}.
The total traffic bit rate in the network is
\begin{eqnarray}
r_T = \left [ (P - H) \left(\frac{N_l}{\alpha} + N_m + \alpha N_h \right)
       + N_r + \alpha H \right] \sigma \bar{L}.
\end{eqnarray}
For notational convenience, we define
the equivalent number of medium bit rate traffic nodes as
\begin{eqnarray}
\eta_{\alpha} = (P - H) \left(\frac{N_l}{\alpha} + N_m + \alpha N_h \right)
       + N_r + \alpha H,
\end{eqnarray}
with which we express the medium traffic bit rate in terms of
$r_T$ as $\sigma \bar{L} = r_T / \eta_{\alpha}$.

For the upstream WDM channels, we note that an LR ONU
sends a packet with probability $1 - \beta$ over the
upstream WDM channels. Hence,
\begin{eqnarray}
\lambda^{W,u} = \frac{r_T}{\eta_{\alpha}}
     \left[\frac{N_l}{\alpha} + N_m + \alpha N_h (1 - \beta)  \right]
\end{eqnarray}
and
\begin{eqnarray}  \label{rt_WDMupnunu:eqn}
r_T < \frac{\eta_{\alpha} (W + 1) C_W}
       {\frac{N_l}{\alpha} + N_m + \alpha N_h (1-\beta)}.
\end{eqnarray}
for our scenario for $\alpha = 2$ and $\eta_{\alpha} = 102$, we obtain
for $\beta = 0.24$: $r_T < 32.59943 C_W$,
for $\beta = 0.5$: $r_T < 38.25 C_W$, and
for $\beta = 1$: $r_T < 57.375 C_W$.

For the WDM downstream channels when using signal reflection for
upstream transmissions, traffic contributions are made by
$(i)$ the transmissions from the WDM ONUs at the considered EPON to the other
WDM ONUs at the EPON,
$(ii)$ the transmissions by all other WDM ONUs
and the ring nodes in the network to all the ONUs in the considered EPON,
$(iii)$ the transmissions by all the LR ONUs and the remote CO to the
WDM ONUs in the considered EPON giving for (\ref{lambda_WDM_down:eqn}):
\begin{eqnarray}
\lambda^{W,d} &=& \frac{r_T}{\eta_{\alpha}}
 \left[\left(\frac{N_l}{\alpha} + N_m \right) \frac{N_l+ N_m - 1}{\eta-1}
   \right.  \\
&&\left. + \left((P-H) \left(\frac{N_l}{\alpha} + N_m \right) + N_r \right)
  \frac{N}{\eta-1} \right. \nonumber \\
 && \left. +  \alpha ( (P-H) N_h + H ) (1 - \beta)
  \frac{N_l+N_m}{\eta_{TWr}}  \right], \nonumber
\end{eqnarray}
Defining for the number of nodes not
connected to the AWG
$\eta_{TWr} = (P-H) (N_l + N_m) + N_r$ and
$\eta_{TWR\alpha} = (P-H) (\frac{N_l}{\alpha} + N_m) + N_r$
gives the limit
\begin{eqnarray}  \label{rt_WDMdownupnunu:eqn}
r_T < \frac{\eta_{\alpha} (W+1)C_W}
   {  \frac{\left(\frac{N_l}{\alpha} + N_m \right) (N_l+ N_m - 1)}{\eta-1}
   +   \frac{\eta_{TWr\alpha} N}{\eta - 1}
     +  \frac{\alpha (1 - \beta) \eta_{LH}(N_l + N_m)}{\eta_{TWr}}}.
\end{eqnarray}
The resulting throughput limits are
$r_T < 28.4$ Gbps for $\beta = 0.24$,
$r_T < 32.5$ Gbps for $\beta = 0.5$, and
$r_T < 45.2$ Gbps for $\beta = 1$.

For the AWG we obtain from (\ref{lambdaA:eqn}) the highest load
on the AWG channels interconnecting two EPONs as
\begin{eqnarray}
\lambda^A = \frac{r_T}{\eta_{\alpha}} \alpha N_h \beta \frac{N_h}{\eta_{LH}}
\end{eqnarray}
and the corresponding limit
\begin{eqnarray}   \label{rt_AWGupnunu:eqn}
r_T < \frac{\eta_{\alpha} \eta_{LH} C_A}
        {\alpha \beta N_h^2}.
\end{eqnarray}
For our scenario, we obtain
for $\beta = 1$ the aggregate throughput constraint
$r_T < 199.2 \mbox{Gbps}$, indicating
that the AWG channels are still utilized to less than 25\%.

\section{Conclusions}
\label{sec:conclusions}
We have developed a comprehensive probabilistic analysis for
evaluating the packet throughput-delay performance of next-generation
PONs (NG-PONs).
Our analysis accommodates both EPONs and GPONs with their various
next-generation upgrades, as well as a variety of all-optical
interconnections of NG-PONs.
Our numerical results illustrate the use of our analysis to
evaluate the throughput delay performance of upgrades that increase the
transmission line rates or wavelength counts.
We also demonstrate the identification of network bottlenecks using
our analysis.

\vspace{\baselineskip}

\begin{centering}
\textsc{Appendix}

\end{centering}

\begin{centering}
\textsc{Specific Capacity Limits for Uniform and Non-uniform Traffic}

\end{centering}

In this appendix we consider an NG-PON with three ring nodes
between two OLTs/COs, whereas only one ring node is considered in
Section~\ref{num:sec}. Further, we consider in this appendix
low traffic ONUs to be TDM ONUs, medium traffic ONUs to be WDM
ONUs and high traffic ONUs to be LR ONUs; whereas, in 
Section~\ref{num:sec} both low and medium traffic ONUs are
considered WDM ONUs and high traffic ONUs are considered LR ONUs.

\textit{A. Uniform Source-Uniform Destination Traffic}
We initially consider uniform source traffic where all nodes generate the same
traffic bit rate, i.e.,
$\bar{L} \sigma(i) = \bar{L} \sigma \ \forall i \in \mathcal{N}$,
in conjunction with uniform destination traffic where a packet generated
at a node $i$ is destined to any of the other nodes with equal probability,
i.e., $r(i,j) = r_i / (\eta - 1) \ \forall j \neq i$.
Note that for the uniform source traffic,
\begin{eqnarray}
r_T = \eta \bar{L} \sigma.
\end{eqnarray}

We consider EPON configurations with
(a) $N_T = 32$,
(b) $N_T = 24$, $N_W = N_L = 4$,
(c) $N_T = 16$, $N_W = N_L = 8$, and
(d) $N_T = 4$, $N_W = N_L = 14$.
A given LR ONU/hotspot CO $i$
sends all traffic destined to other LR ONUs/the hotspot CO $j$
over the AWG subnetwork, i.e.,
$r^A (i) = [(P-2) N_L + N_L - 1 + 1] \sigma / (\eta - 1)$, since
there are $N_L$ LR ONUs at each of the other COs,
$N_L - 1$ other LR ONUs at the CO that the considered
LR ONU $i$ is attached to, and one hotspot.
If transmission over the AWG is possible,
the corresponding traffic rates $T(i,j)$ and $r(i,j)$ for traffic over the
EPON and PSC/ring subnetwork are set to zero. That is,
the total traffic generation rate at each node is constant, and the
generated traffic is either sent over the EPON and PSC/ring subnetwork,
or, if possible, over the AWG subnetwork.

From the stability conditions in Section~\ref{analysis:sec}, we readily
see that for the TDM EPON channel the
upstream condition (\ref{eqn:tdmcap})
gives $\lambda^{T,u.k} = \frac{r_T}{\eta} N_T$ since
each node generates a bit rate of $\frac{r_T}{\eta}$ and the $N_T$
TDM nodes send all their traffic upstream on the TDM channel.
Hence, the stability condition in terms of $r_T$ takes the form
\begin{eqnarray} \label{rt_TDMup:eqn}
r_T < \frac{\eta C}{ N_T}.
\end{eqnarray}

For the WDM EPON channel upstream condition (\ref{eqn:wdmcap}),
we note that the $N_W$ WDM ONUs send all their upstream traffic and the
$N_L$ LR ONUs send all traffic not destined to other LR ONUs/hotspot,
i.e., all traffic destined to TDM/WDM ONUs and ring nodes,
upstream on the WDM channels (or the TDM channel).
Denoting for notational convenience the number of
TDM/WDM ONUs and ring nodes by
\begin{eqnarray}
\eta_{TWr} = (P - H) (N_T + N_W) + N_r,
\end{eqnarray}
we obtain
$\lambda^{W,u.k} = \frac{r_T}{\eta}
 ( N_W + N_L \frac{\eta_{TWr}}{\eta - 1})$,
resulting in the condition
\begin{eqnarray} \label{rt_WDMup:eqn}
r_T < \frac{\eta (\eta - 1) (W + 1) C}
    {(\eta-1) (N_T + N_W) + \eta_{TWr} N_L}.
\end{eqnarray}

For the downstream TDM channel, we note that
the $N_T$ TDM nodes at the considered EPON send a packet to one of the
other $N_T-1$ TDM nodes in the same EPON with probability
$(N_T - 1) / (\eta - 1)$. The other $\eta - N_T$ nodes in the network
send a packet to one of the $N_T$ nodes in the considered EPON with
probability $N_T / (\eta - 1)$.
Thus,
\begin{eqnarray}  \lambda^{T,d,k} = \frac{r_T}{\eta}
\left[ N_T \cdot \frac{N_T - 1}{ \eta - 1}
        + (\eta- N_T) \frac{N_T }{ \eta - 1} \right],
\end{eqnarray}
which simplifies to the stability condition (\ref{rt_TDMup:eqn}).

For the downstream WDM channel condition for
reflection of the downstream signal (\ref{lambda_WDM_down:eqn})
there are three contributors to downstream WDM channel traffic:
$(i)$ the $N_W$ WDM ONUs at the considered EPON sending to other WDM
ONUs at this EPON contributing
$\frac{r_T}{\eta} N_W \frac{N_W -1}{\eta-1}$,
$(ii)$ all other nodes in the network sending to the $N_W$ WDM ONUs,
 contributing $\frac{r_T}{\eta} (\eta - N_W) \frac{N_W}{\eta - 1}$,
and
$(iii)$ all TDM/WDM ONUs and ring nodes sending to the
LR ONUs in the considered EPON contributing
$\frac{r_T}{\eta} \eta_{TWr} \frac{N_L}{\eta - 1}$,
resulting in
\begin{eqnarray}  \label{rt_WDMdown:eqn}
\lambda^{W,d,k} = \frac{r_T}{\eta (\eta - 1)}
     \left[ (\eta - 1) N_W  + \eta_{TWr} N_L \right].
\end{eqnarray}
Inserting into (\ref{lambda_WDM_down:eqn}) results in a condition identical
to (\ref{rt_WDMup:eqn}).

For upstream transmission with an empty carrier, condition
(\ref{lambda_WDM_down_empty:eqn}) gives
\begin{eqnarray}   \label{rt_WDMempty:eqn}
r_T < \frac{\eta (\eta - 1) (W+1) C}{2 [ (\eta - 1) (N_T + N_W)
             + \eta_{TWr} N_L ] }.
\end{eqnarray}

For analysis of the ring/PSC stability condition
(\ref{lambdaPl:eqn}), we fix the ring network to the structure
illustrated in Fig.~\ref{fig:AWG} with $N_r = 12$ ring nodes, $P
= 4$ COs (of which $H = 1$ is a hotspot CO), and with three ring nodes
between two COs. The highest traffic rate on a PSC channel arises
for the uniform traffic pattern on the channel toward a CO with
attached EPON when the hotspot CO is four ring hops from the
considered CO. We refer henceforth to the considered CO as the
``target CO'', the CO opposite of the target CO around the ring as
the ``opposite CO'', and the CO situated four hops along the ring
from the target CO as the ``adjacent CO'' (which is located opposite
the hotspot CO). We have the following contributions to the traffic
load on the PSC home channel of the target CO:
\begin{itemize}
\item The opposite CO has
$N_T + N_W$ TDM/WDM ONUs sending
$(i)$ to the $N$ nodes in the EPON attached to the target CO, plus
$(ii)$ to the two ring nodes adjacent to the target CO, plus
$(iii)$ with probability one half to the
two ring nodes situated two hops
from the target CO over the target home channel.
In addition, the opposite CO has $N_L$ LR ONUs
sending
$(i)$ to the $N_T + N_W$ TDM/WDM ONUs in the EPON attached
to the target CO, plus
$(ii)$ to the two ring nodes adjacent to the target CO, plus
$(iii)$ with probability one half to the
two ring nodes situated two hops
from the target CO over the target channel.
Thus, the opposite CO contributes
\begin{eqnarray}
\frac{r_T}{\eta} (N_T + N_W) \frac{N + 2 + \frac{1}{2} 2}{\eta-1} +
   \frac{r_T}{\eta} N_L \frac{N_T + N_W + 2 + \frac{1}{2} 2}{\eta-1}.
\end{eqnarray}
\item The adjacent CO has the same contribution as the opposite CO,
except that only the traffic to the two ring nodes adjacent to the
target CO is sent over the PSC; the ring nodes situated
two hops from the target CO are reached with two hops over the ring
(versus three hops over the PSC and then ring).
Furthermore, the hotspot contributes as much as one
LR ONU at the adjacent CO.
Overall, the contribution from the adjacent CO and hotspot is thus
\begin{eqnarray}
\frac{r_T}{\eta} (N_T + N_W) \frac{N + 2}{\eta-1} +
   \frac{r_T}{\eta} (N_L+1) \frac{N_T + N_W + 2}{\eta-1}.
\end{eqnarray}
\item The four ring nodes situated within two ring hops from the
target CO do not send traffic toward the target CO over the PSC.
\item The two ring nodes situated three hops in either ring direction
from the target CO send to all ONUs attached to the target CO over
the PSC (and directly over the ring to the ring nodes one and two hops
from the target CO) contributing
\begin{eqnarray}
\frac{r_T}{\eta} 2 \frac{N}{\eta - 1}.
\end{eqnarray}
\item The four ring nodes situated five or six hops in either ring direction
from the target CO send to all ONUs attached to the target CO plus to the
two ring nodes adjacent to the target CO over
the PSC contributing
\begin{eqnarray}
\frac{r_T}{\eta} 4 \frac{N + 2}{\eta - 1}.
\end{eqnarray}
\item The two ring nodes situated seven hops in either ring direction
from the target CO send to all ONUs attached to the target CO, plus to the
two ring nodes adjacent to the target CO, plus with probability one half to
the two ring nodes situated two hops from the target CO over
the PSC contributing
\begin{eqnarray}
\frac{r_T}{\eta} 2 \frac{N + 2 + \frac{1}{2} 2}{\eta - 1}.
\end{eqnarray}
\end{itemize}
Combining these contributions results in the constraint
\begin{eqnarray} \label{rt_PSC:eqn}
r_T < \frac{\eta (\eta - 1) C}{(N+3)(N_T +N_W+2) + (N+2)(N_T + N_W+4)
             + N_L (2N_T + 2N_W + 5) + 2 N + N_T + N_W + 2}.
\end{eqnarray}

For the considered uniform traffic,
the maximum $\lambda^A (k,l)$ arises for the AWG channel from
a CO with an attached EPON to another CO with an attached EPON.
Specifically, the tightest AWG constraint (\ref{lambdaA:eqn}) becomes
\begin{eqnarray} \label{rt_AWG:eqn}
r_T <  \frac{\eta (\eta - 1) c C }{N_L^2 }.
\end{eqnarray}

\begin{table}
\caption{Summary of stability conditions and resulting limits on
total traffic bit rate in network $r_T$ for different uniform
source-uniform destination traffic scenarios}
\label{usud:tab}
\begin{tabular}{llrrrr}
Cap. Const. & $r_T$ lim.\ &  $N_T = 32$ & $N_T = 24$, $N_W = N_L = 4$ &
      $N_T = 16$, $N_W = N_L = 8$  &  $N_T = 4$, $N_W = N_L = 14$ \\
T,u (\ref{eqn:tdmcap})  & (\ref{rt_TDMup:eqn})                                                                   &     3.41   &   4.54    &    6.81     &  27.25  \\
W,u (\ref{eqn:wdmcap})  & (\ref{rt_WDMup:eqn})                                                               &    N/A        &   6.91    &    7.21    &  8.21  \\
T,d (\ref{lambda_TDM_down:eqn})  & (\ref{rt_TDMup:eqn})                                             &     3.41   &   4.54    &    6.81     &  27.25  \\
W,d, refl.\ (\ref{lambda_WDM_down:eqn}) & (\ref{rt_WDMup:eqn})                                  &      N/A      &   6.91    &    7.21    &  8.21  \\
W,d, empty car.\ (\ref{lambda_WDM_down_empty:eqn}) & (\ref{rt_WDMempty:eqn})    &       N/A       &    3.45   &    3.61    &  4.10  \\
Ring/PSC (\ref{lambdaPl:eqn}) &  (\ref{rt_PSC:eqn})                                                      &      4.69   &   4.75    &    4.95    &  5.59  \\
AWG (\ref{lambdaA:eqn}) & (\ref{rt_AWG:eqn})                                                                     &     N/A       &    735.75    &    183.94   &  60.06  \\
\end{tabular}
\end{table}

\textit{B. Non-uniform Source-Uniform Destination Traffic}
In this section we consider non-uniform source traffic whereby
nodes generate different traffic rates.
However, we continue to consider uniform destination traffic, i.e.,
each generated packet is destined to any of the other $\eta - 1$ nodes with
equal probability.
More specifically, $N_m$ of the ONUs in each EPON, as well as
all ring nodes generate a medium traffic bit rate $\sigma \bar{L}$.
Furthermore, given a source traffic non-uniformity $\alpha,\ \alpha \geq 1$,
$N_l$ ONUs in each EPON generate
a low traffic bit rate $\sigma \bar{L}/\alpha$.
Also, $N_h$ ONUs in each EPON and the hotspot CO generate
a high traffic bit rate $\alpha \sigma \bar{L}$.
The total traffic bit rate in the network is
\begin{eqnarray}
r_T = \left [ (P - H) \left(\frac{N_l}{\alpha} + N_m + \alpha N_h \right)
       + N_r + \alpha H \right] \sigma \bar{L}.
\end{eqnarray}
For notational convenience, we define
the equivalent number of medium bit rate traffic nodes as
\begin{eqnarray}
\eta_{\alpha} = (P - H) \left(\frac{N_l}{\alpha} + N_m + \alpha N_h \right)
       + N_r + \alpha H,
\end{eqnarray}
with which we express the medium traffic bit rate in terms of
$r_T$ as $\sigma \bar{L} = r_T / \eta_{\alpha}$.
The traffic routing rules from the preceding section apply, that is,
traffic is sent over the AWG whenever possible.

Initially, all ONUs are TDM ONUs.
Then, the medium rate ONUs are upgraded to WDM ONUs and the high rate ONUs
are upgraded to LR ONUs.
For the scenario with only TDM ONUs, the upstream TDM EPON channel
condition (\ref{eqn:tdmcap})
gives $\lambda^{T,u.k} = \frac{r_T}{\eta_{\alpha}} \cdot
[\frac{N_l}{\alpha} + N_m + \alpha N_h]$,
giving the stability condition in terms of $r_T$
\begin{eqnarray} \label{rt_TDMupa:eqn}
r_T < \frac{\eta_{\alpha} C}{\frac{N_l}{\alpha} + N_m + \alpha N_h}.
\end{eqnarray}
For the downstream TDM channel condition (\ref{lambda_TDM_down:eqn})
in the initial (non-upgraded) scenario,
there is a load contribution
$\frac{r_T}{\eta_{\alpha}} \left( \frac{N_l}{\alpha} + N_m + \alpha N_h \right)
\frac{N - 1}{\eta-1}$
due to the ONUs in the considered EPON sending to other ONUs in the same
EPON, and a contribution
$\frac{r_T}{\eta_{\alpha}} [ (P - H - 1)
  \left( \frac{N_l}{\alpha} + N_m + \alpha N_h \right) + N_r + \alpha H ]
\frac{N}{\eta-1} =
\frac{r_T}{\eta_{\alpha}}
  [ \eta_{\alpha} - \left( \frac{N_l}{\alpha} + N_m + \alpha N_h \right) ]
\frac{N}{\eta-1}$
due to the other network nodes sending to the ONUs in the considered EPON,
resulting in the limit
\begin{eqnarray}  \label{rt_TDMdowna:eqn}
r_T < \frac{\eta_{\alpha} (\eta - 1) C}
  {\eta_{\alpha} N - \left( \frac{N_l}{\alpha} + N_m + \alpha N_h \right) }.
\end{eqnarray}
For the PSC in the initial setting, we obtain following the
analysis above the contributions:
\begin{itemize}
\item Opposite CO: $\frac{r_T}{\eta_{\alpha}}
    \left( \frac{N_l}{\alpha} + N_m + \alpha N_h \right)
       \frac{N + 2 + \frac{1}{2} 2}{\eta - 1}$
\item Adjacent CO and hotspot:
    $\frac{r_T}{\eta_{\alpha}}
    \left( \frac{N_l}{\alpha} + N_m + \alpha (N_h + 1) \right)
       \frac{N + 2 }{\eta - 1}$
\item Ring nodes three hops from target CO:
   $\frac{r_T}{\eta_{\alpha}} 2 \frac{N}{\eta-1}$
\item Ring nodes four and five hops from target CO:
   $\frac{r_T}{\eta_{\alpha}} 4 \frac{N + 2}{\eta-1}$
\item Ring nodes six hops from target CO:
   $\frac{r_T}{\eta_{\alpha}} 2 \frac{N + 2 + \frac{1}{2}2}{\eta-1}$
\end{itemize}
The resulting stability limit is
\begin{eqnarray}  \label{rt_PSCa:eqn}
r_T < \frac{\eta_{\alpha} (\eta-1) C}
   {\left( \frac{N_l}{\alpha} + N_m + \alpha N_h \right) (2N + 5) +
            8 N + 14 + \alpha(N+2)}.
\end{eqnarray}

Next, for the upgraded scenario,
$N_T = N_l$, $N_W = N_m$, and $N_L = N_h$.
For the upstream TDM channel,
\begin{eqnarray}  \label{rt_TDMupaup:eqn}
r_T < \frac{\eta_{\alpha} C}{\frac{N_l}{\alpha}}.
\end{eqnarray}
For the upstream WDM channels we obtain
\begin{eqnarray}
\lambda^{W,u,k} = \frac{r_T}{\eta_{\alpha}} \left(N_m +
        \alpha N_h \frac{\eta_{TWr}}{\eta-1} \right)
\end{eqnarray}
and the corresponding limit
\begin{eqnarray}  \label{rt_WDMupaup:eqn}
r_T < \frac{\eta_{\alpha} (\eta-1) (W + 1) C}
       {(\eta-1) \left( \frac{N_l}{\alpha} + N_m \right) + \eta_{TWr}
      \alpha N_L}.
\end{eqnarray}

For the downstream TDM channel we obtain by considering the transmissions from
the each of the $N_l$ TDM ONUs in the considered EPON
sending with bit rate $r_T/(\eta_{\alpha} \alpha)$ and a packet being destined
to the other
$N_l  - 1$ TDM ONUs in the EPON with probability
$(N_l - 1)/(\eta-1)$ the load contribution
$\frac{r_T}{\eta_{\alpha}} \frac{N_l}{\alpha} \frac{N_l - 1}{\eta - 1}$.
Further, considering the transmissions from
all other nodes generating the traffic bit rate
$\frac{r_T}{\eta_{\alpha}}
  \left[ (P - H) \left(\frac{N_l}{\alpha} + N_m + \alpha N_h \right)
       + N_r + \alpha H - \frac{N_l}{\alpha} \right] =
   \frac{r_T}{\eta_{\alpha}} \left[ \eta_{\alpha} - \frac{N_l}{\alpha} \right]$
of which the fraction $N_l/(\eta-1)$ is destined to the TDM ONUs in the
considered EPON.
Hence,
\begin{eqnarray}
\lambda^{T,d,k} = \frac{r_T}{\eta_{\alpha}}
   \left[ \frac{N_l}{\alpha} \frac{N_l - 1}{\eta - 1} +
           \left( \eta_{\alpha} - \frac{N_l}{\alpha} \right)
           \frac{N_l}{\eta-1} \right],
  \end{eqnarray}
resulting in the limit
\begin{eqnarray}  \label{rt_TDMdownaup:eqn}
r_T < \frac{\eta_{\alpha} (\eta-1)C}{\frac{N_l}{\alpha}
          \left( \alpha \eta_{\alpha}  -1 \right)}.
\end{eqnarray}

For the WDM downstream channels when using signal reflection for
upstream transmissions, we denote for convenience
\begin{eqnarray}
\eta_{TWr\alpha} = (P-H) \left( \frac{N_l}{\alpha} + N_m \right) +
  N_r
\end{eqnarray}
and obtain
by considering $(i)$ the transmissions from the $N_m$ medium bit rate
ONUs at the EPON to the other medium bit rates ONUs at the EPON,
$(ii)$ the transmissions by all other nodes in the networks to the
medium bit rate ONUs in the EPON, and
$(iii)$ the transmissions by all TDM/WDM ONUs and ring nodes in the network
to the high bit rate ONUs in the EPON
\begin{eqnarray}
\lambda^{W,d,k} = \frac{r_T}{\eta_{\alpha}}
 \left[ N_m \frac{N_m - 1}{\eta-1} + (\eta_{\alpha} - N_m) \frac{N_m}{\eta-1}
       + \eta_{TWr\alpha} \frac{N_h}{\eta-1} \right],
\end{eqnarray}
resulting in the limit
\begin{eqnarray}  \label{rt_WDMdownaup:eqn}
r_T < \frac{\eta_{\alpha} (\eta-1) (W+1)C}
   {  \left( \alpha \eta_{\alpha} -1 \right) \frac{N_l}{\alpha}
      + (\eta_{\alpha} - 1) N_m +  \eta_{TWr\alpha} N_h }.
\end{eqnarray}

For the WDM downstream channels using an empty carrier for upstream
transmissions, we obtain
\begin{eqnarray}  \label{rt_WDMemptyaup:eqn}
r_T < \frac{\eta_{\alpha} (\eta-1) (W+1)C }
  { \eta_{\alpha} (\eta-1) N_l + (\eta_{\alpha} + \eta -2) N_m +
     ( \alpha \eta_{TWr} + \eta_{TWr\alpha}) N_h }.
\end{eqnarray}

For the PSC, we have the contributions:
\begin{itemize}
\item Opposite CO: $\frac{r_T}{\eta_{\alpha}}
    \left( \frac{N_l}{\alpha} + N_m \right)
       \frac{N + 3}{\eta - 1}
         + \frac{r_T}{\eta_{\alpha}}  \alpha N_h
           \frac{N_l + N_m + 3}{\eta-1}$
\item Adjacent CO and hotspot:
    $\frac{r_T}{\eta_{\alpha}}
    \left( \frac{N_l}{\alpha} + N_m \right)
       \frac{N + 2 }{\eta - 1}
          + \frac{r_T}{\eta_{\alpha}} \alpha (N_h + 1)
                          \frac{N_l + N_m + 2}{\eta-1}$
\item Ring nodes three hops from target CO:
   $\frac{r_T}{\eta_{\alpha}} 2 \frac{N}{\eta-1}$
\item Ring nodes four and five hops from target CO:
   $\frac{r_T}{\eta_{\alpha}} 4 \frac{N + 2}{\eta-1}$
\item Ring nodes six hops from target CO:
   $\frac{r_T}{\eta_{\alpha}} 2 \frac{N + 3}{\eta-1}$
\end{itemize}
The resulting stability limit is
\begin{eqnarray}  \label{rt_PSCaup:eqn}
r_T < \frac{\eta_{\alpha} (\eta-1) C}
   { (N+3) \left( \frac{N_l}{\alpha} + N_m + 2 \right) +
     (N+2) \left( \frac{N_l}{\alpha} + N_m + 4 \right) +
     \alpha N_h (2 N_l + 2 N_m + 5) + 2 N + N_l + N_m + 2}.
\end{eqnarray}

For the AWG we obtain the highest load
$\frac{r_T}{\eta_{\alpha}} \alpha N_h \frac{N_h}{\eta-1}$
and the corresponding limit
\begin{eqnarray}   \label{rt_AWGaup:eqn}
r_T < \frac{\eta_{\alpha} (\eta-1) c C}
        {\alpha N_h^2}.
\end{eqnarray}

We initially consider $N_l = 16$, $N_m = 8$, $N_h = 8$ with $\alpha = 2$.

\begin{table}
\caption{Summary of stability conditions and resulting limits on
total traffic bit rate in network $r_T$ for different nonuniform
source-uniform destination traffic scenarios for the TDM ONUs-only
(non-upgraded network and the upgraded network)}
\label{nusud:tab}
\begin{tabular}{llrrrr}
Cap. Const. & $r_T$ lim.\ &   &  &    &  \\
\multicolumn{6}{c}{TDM ONUs only}  \\
T,u (\ref{eqn:tdmcap})  & (\ref{rt_TDMupa:eqn})   &      3.44      &       &        &    \\
T,d (\ref{lambda_TDM_down:eqn})  & (\ref{rt_TDMdowna:eqn})      &      3.41      &       &        &    \\
Ring/PSC (\ref{lambdaPl:eqn}) &  (\ref{rt_PSCa:eqn})    &    4.67        &       &        &    \\
\multicolumn{6}{c}{Upgraded Network}  \\
T,u (\ref{eqn:tdmcap})  & (\ref{rt_TDMupaup:eqn})   &    13.75
        &       &        &    \\
W,u (\ref{eqn:wdmcap})  & (\ref{rt_WDMupaup:eqn})  &     7.73       &       &        &    \\
T,d (\ref{lambda_TDM_down:eqn})  & (\ref{rt_TDMdownaup:eqn})      &     6.78       &       &        &    \\
W,d, refl.\ (\ref{lambda_WDM_down:eqn}) & (\ref{rt_WDMdownaup:eqn})      &   7.65         &       &        &    \\
W,d, empty car.\ (\ref{lambda_WDM_down_empty:eqn}) &
  (\ref{rt_WDMemptyaup:eqn})     &      0.12      &       &        &    \\
Ring/PSC (\ref{lambdaPl:eqn}) &  (\ref{rt_PSCaup:eqn})    &    5.28        &       &        &    \\
AWG (\ref{lambdaA:eqn}) & (\ref{rt_AWGaup:eqn})  &     92.81       &       &        &    \\
\end{tabular}
\end{table}

\textit{C. Non-uniform Source-Non-uniform Destination Traffic}
In this section we further build on the non-uniform source traffic
from the preceding section in that we introduce non-uniform destinations.
Specifically, we denote
\begin{eqnarray}
\eta_{LH} = (P-H) N_L + H
\end{eqnarray}
for the total number of LR ONUs and hotspots in the network.
With non-uniform destination traffic,
a packet generated at an LR ONU or hotspot CO is
destined to another LR ONUs/hotspot CO with probability
$\beta,\ (\eta_{LH} - 1)/ (\eta - 1) \leq \beta \leq 1$.
A packet that is destined to another LR ONUs/hotspot CO is destined to
any of the $\eta_{LH} - 1$ other LR ONUs/hotspot COs
with equal probability.
Packets generated at ring nodes, WDM or TDM ONUs are
destined to any of the other network nodes with equal probability.
We consider in the following the upgraded network scenario with
$N_T = N_l$, $N_W = N_m$, and $N_L = N_h$.
The upstream TDM channel carries only traffic from the TDM nodes,
and is therefore not affected by the non-uniformity of the traffic
destinations. Hence, the condition (\ref{rt_TDMupaup:eqn}) still holds.

For the upstream WDM channels, we note that an LR ONU
sends a packet with probability $1 - \beta$ over the
upstream WDM channels. Hence,
\begin{eqnarray}
\lambda^{W,u,k} = \frac{r_T}{\eta_{\alpha}}
     \left[N_m + \alpha N_h (1 - \beta)  \right]
\end{eqnarray}
and
\begin{eqnarray}  \label{rt_WDMupnunu:eqn}
r_T < \frac{\eta_{\alpha} (W + 1) C}
       {\frac{N_l}{\alpha} + N_m + \alpha N_h (1-\beta)}.
\end{eqnarray}

For the downstream TDM channel, the TDM ONUs at the considered EPON
make the load contribution
$\frac{r_T}{\eta_{\alpha}} \frac{N_l}{\alpha} \frac{N_l - 1}{\eta - 1}$,
as above for the analysis leading to (\ref{rt_TDMdownaup:eqn}).
All other TDM ONUs together with all WDM ONUs and ring nodes make the
contribution
$\frac{r_T}{\eta_{\alpha}}
  \left[ (P - H) \left(\frac{N_l}{\alpha} + N_m \right)
                                   + N_r - \frac{N_l}{\alpha}
      \right]  \frac{N_l}{\eta - 1} =
   \frac{r_T}{\eta_{\alpha}}
           \left[ \eta_{TWr\alpha} - \frac{N_l}{\alpha} \right]
    \frac{N_l}{\eta-1}$.
Furthermore, the LR ONUs and hotspots make the contribution
$\frac{r_T}{\eta_{\alpha}}
  \left( (P - H) \alpha N_h + \alpha H \right) (1- \beta)
      \frac{N_l}{\eta_{TWr}}
 = \frac{r_T}{\eta_{\alpha}}
  \alpha \eta_{LH} (1- \beta)
      \frac{N_l}{\eta_{TWr}}$,
resulting in
\begin{eqnarray}
\lambda^{T,d,k} = \frac{r_T N_l}{\eta_{\alpha}}
   \left[ \frac{1}{\eta - 1} \left( \eta_{TWr\alpha} - \frac{1}{\alpha} \right) +
          \frac{\alpha (1 - \beta)}{\eta_{TWr}} \eta_{LH} \right],
  \end{eqnarray}
and the limit
\begin{eqnarray}  \label{rt_TDMdownupnunu:eqn}
r_T < \frac{\eta_{\alpha} C}
   {N_l \left\{ \frac{1}{\eta - 1}
            \left( \eta_{TWr\alpha} - \frac{1}{\alpha}\right)
      +  \frac{\alpha (1 - \beta)}{\eta_{TWr}} \eta_{LH}
           \right\} }.
\end{eqnarray}

For the WDM downstream channels when using signal reflection for
upstream transmissions, traffic contributions are made by
$(i)$ the transmissions from the $N_m$ medium bit rate
ONUs at the EPON to the other medium bit rates ONUs at the EPON,
$(ii)$ the transmissions by all other TDM/WDM ONUs
and ring nodes in the networks to the
medium bit rate ONUs in the EPON,
$(iii)$ the transmissions by the LR ONUs and hotspots to the medium
bit rate ONUs in the EPON, and
$(iv)$ the transmissions by all TDM/WDM ONUs and ring nodes in the network
to the high bit rate ONUs in the EPON
\begin{eqnarray}
\lambda^{W,d,k} = \frac{r_T}{\eta_{\alpha}}
 \left[\frac{N_m}{\eta-1} (\eta_{TWr\alpha} - 1)
      +  \alpha (1-\beta) \eta_{LH} \frac{N_m}{\eta_{TWr}}
      + \eta_{TWr\alpha} \frac{N_h}{\eta-1} \right],
\end{eqnarray}
resulting in the limit
\begin{eqnarray}  \label{rt_WDMdownupnunu:eqn}
r_T < \frac{\eta_{\alpha} (W+1)C}
   { ( \eta_{TWr\alpha} -1) \frac{N_m}{\eta - 1}
     + \alpha (1 - \beta) \eta_{LH} \frac{N_m}{\eta_{TWr}}
      + \eta_{TWr\alpha} \frac{N_h}{\eta - 1} }.
\end{eqnarray}

For the WDM downstream channels using an empty carrier for upstream
transmissions, we obtain
\begin{eqnarray}  \label{rt_WDMemptyupnunu:eqn}
r_T < \frac{\eta_{\alpha} (W+1)C }
  { \frac{N_l}{\alpha} + N_m + \alpha (1 - \beta) N_h
   + \frac{N_l}{\eta - 1} \left( \eta_{TWr\alpha} - \frac{1}{\alpha} \right)
   + \frac{N_m}{\eta - 1} (\eta_{TWr\alpha} - 1)
  + \frac{\alpha (1 - \beta) \eta_{LH}}{\eta_{TWr}} (N_l + N_m)
  + \eta_{TWr\alpha} \frac{N_h}{\eta - 1}}.
\end{eqnarray}

For the PSC, we have the contributions:
\begin{itemize}
\item Opposite CO: $\frac{r_T}{\eta_{\alpha}}
    \left( \frac{N_l}{\alpha} + N_m \right)
       \frac{N + 3}{\eta - 1}
         + \frac{r_T}{\eta_{\alpha}}  \alpha N_h (1 - \beta)
           \frac{N_l + N_m + 3}{\eta_{TWr}}$
\item Adjacent CO and hotspot:
    $\frac{r_T}{\eta_{\alpha}}
    \left( \frac{N_l}{\alpha} + N_m \right)
       \frac{N + 2 }{\eta - 1}
          + \frac{r_T}{\eta_{\alpha}} \alpha (N_h + 1) (1 - \beta)
                          \frac{N_l + N_m + 2}{\eta_{TWr}}$
\item Ring nodes three hops from target CO:
   $\frac{r_T}{\eta_{\alpha}} 2 \frac{N}{\eta-1}$
\item Ring nodes four and five hops from target CO:
   $\frac{r_T}{\eta_{\alpha}} 4 \frac{N + 2}{\eta-1}$
\item Ring nodes six hops from target CO:
   $\frac{r_T}{\eta_{\alpha}} 2 \frac{N + 3}{\eta-1}$
\end{itemize}
The resulting stability limit is
\begin{eqnarray}  \label{rt_PSCupnunu:eqn}
r_T < \frac{\eta_{\alpha}  C}
   { \frac{1}{\eta - 1} \left[ \left( \frac{N_l}{\alpha} + N_m \right)
      (2N + 5) + (8N + 14) \right] +
     \frac{\alpha (1-\beta)}{\eta_{TWr}}
             [ N_h (N_l + N_m + 3) + (N_h + 1) (N_l + N_m + 2)]}.
\end{eqnarray}

For the AWG we obtain the highest load
$\frac{r_T}{\eta_{\alpha}} \alpha N_h \beta \frac{N_h}{\eta_{LH}}$
and the corresponding limit
\begin{eqnarray}   \label{rt_AWGupnunu:eqn}
r_T < \frac{\eta_{\alpha} \eta_{LH} c C}
        {\alpha \beta N_h^2}.
\end{eqnarray}

We initially consider $N_l = 16$, $N_m = 8$, $N_h = 8$
with $\alpha = 2$ and $\beta=0.75$.

\begin{table}
\caption{Summary of stability conditions and resulting limits on
total traffic bit rate in network $r_T$ for different nonuniform
source-nonuniform destination traffic scenarios)}
\label{nusnud:tab}
\begin{tabular}{llrrrr}
Cap. Const. & $r_T$ lim.\ &   &  &    &  \\
T,u (\ref{eqn:tdmcap})  & (\ref{rt_TDMupaup:eqn})   &  13.75        &       &        &    \\
W,u (\ref{eqn:wdmcap})  & (\ref{rt_WDMupnunu:eqn})  &  11             &       &        &    \\
T,d (\ref{lambda_TDM_down:eqn})  & (\ref{rt_TDMdownupnunu:eqn})      &   9.83      &       &        &    \\
W,d, refl.\ (\ref{lambda_WDM_down:eqn}) & (\ref{rt_WDMdownupnunu:eqn})      &   21.99       &       &        &    \\
W,d, empty car.\ (\ref{lambda_WDM_down_empty:eqn}) &
  (\ref{rt_WDMemptyupnunu:eqn})     &    5.34        &       &        &    \\
Ring/PSC (\ref{lambdaPl:eqn}) &  (\ref{rt_PSCupnunu:eqn})    &    7.14         &       &        &    \\
AWG (\ref{lambdaA:eqn}) & (\ref{rt_AWGupnunu:eqn})  &     28.65     &       &        &    \\
\end{tabular}
\end{table}

\subsection{Empty Carrier vs.\ Signal Reflection for WDM Channel Upstream
Transmission}

We initially consider the following elementary per-TDM cycle
switching policy for switching between upstream and downstream
transmission on the $W^k$ WDM channels. For ease of exposition, we
explain the policy first for $W^k = 1$ channel. During a given cycle
on the TDM channel, the OLT collects (a) both the REPORT messages
(upstream transmission requests) from the attached ONUs as well as
(b) the packets arriving at the OLT for forwarding downstream on the
WDM channels to the ONUs. At the end of the cycle, i.e., when REPORT
messages from all attached ONUs have been received, the OLT
schedules: (a) All the requested upstream transmissions to arrive
contiguously (spaced by appropriate guard intervals) at the OLT. We
initially consider Gated service whereby the full upstream
transmission request is granted and schedule the upstream grants in
a first-come-first-served manner. At the end of the upstream
transmissions, a switch over is scheduled. (b) After the switchover,
the OLT schedules the transmission of all the downstream packets
collected in the cycle in first-come-first-served manner, followed
by a switchover. With this per-TDM cycle switching policy the OLT
schedules two switchovers corresponding to each TDM cycle.
Scheduling takes place at the end of every cycle on the TDM channel,
i.e., when again REPORTs from all ONUs have been received. Note that
the length of the upstream plus downstream transmission schedule on
the WDM channel is not necessarily equal to the length of the cycle
on the TDM channel. If a large upstream traffic volume is reported
and a large downstream traffic volume collected the OLT may schedule
the upstream and downstream transmissions way into the future.
However, to every cycle on the TDM channel there corresponds one
upstream-plus-downstream transmission schedule on the WDM channel.

We extend the outlined per-TDM cycle scheduling policy for $W^k =1$
WDM channel to $W^k > 1$ WDM channels as follows.
As with $W^k = 1$, the OLT schedules the upstream and downstream transmissions
once per TDM cycle. The computed upstream plus downstream schedule is assigned
to the $W^k$ channels in round-robin fashion. That is, a given
WDM channel is assigned a schedule every $W^k$ cycles on the TDM channel.

\bibliographystyle{./IEEEtran}

\begin{thebibliography}{10}
\providecommand{\url}[1]{#1}
\csname url@samestyle\endcsname
\providecommand{\newblock}{\relax}
\providecommand{\bibinfo}[2]{#2}
\providecommand{\BIBentrySTDinterwordspacing}{\spaceskip=0pt\relax}
\providecommand{\BIBentryALTinterwordstretchfactor}{4}
\providecommand{\BIBentryALTinterwordspacing}{\spaceskip=\fontdimen2\font plus
\BIBentryALTinterwordstretchfactor\fontdimen3\font minus
  \fontdimen4\font\relax}
\providecommand{\BIBforeignlanguage}[2]{{%
\expandafter\ifx\csname l@#1\endcsname\relax
\typeout{** WARNING: IEEEtran.bst: No hyphenation pattern has been}%
\typeout{** loaded for the language `#1'. Using the pattern for}%
\typeout{** the default language instead.}%
\else
\language=\csname l@#1\endcsname
\fi
#2}}
\providecommand{\BIBdecl}{\relax}
\BIBdecl

\bibitem{AuSRGM11}
F.~Aurzada, M.~Scheutzow, M.~Reisslein, N.~Ghazisaidi, and M.~Maier, ``Capacity
  and delay analysis of next-generation passive optical networks {(NG-PONs)},''
  \emph{IEEE Transactions on Communications}, vol.~59, no.~5, pp. 1378--1388,
  May 2011.

\bibitem{Effe07}
F.~Effenberger, D.~Clearly, O.~Haran, G.~Kramer, R.~D. Li, M.~Oron, and
  T.~Pfeiffer, ``An introduction to {PON} technologies,'' \emph{IEEE
  Communications Magazine}, vol.~45, no.~3, pp. S17--S25, Mar. 2007.

\bibitem{KaSGW07}
L.~G. Kazovsky, W.-T. Shaw, D.~Gutierrez, N.~Cheng, and S.-W. Wong,
  ``Next-generation optical access networks,'' \emph{IEEE/OSA J. of Lightwave
  Tech.}, vol.~25, no.~11, pp. 3428--3442, Nov. 2007.

\bibitem{ZhALEY09}
J.~Zhang, N.~Ansari, Y.~Luo, F.~Effenberger, and F.~Ye, ``Next-generation
  {PONs}: A performance investigation of candidate architectures for
  next-generation access stage 1,'' \emph{IEEE Communications Magazine},
  vol.~47, no.~8, pp. 49--57, Aug. 2009.

\bibitem{Lin08}
R.~Lin, ``Next generation {PON} in emerging networks,'' in \emph{Proc.
  OFC/NFOEC}, San Diego, CA, Feb. 2008.

\bibitem{GrEl08}
K.~Grobe and J.-P. Elbers, ``{PON} in adolescence: From {TDMA} to {WDM-PON},''
  \emph{IEEE Communications Mag.}, vol.~46, no.~1, pp. 26--34, Jan. 2008.

\bibitem{KMP0202}
G.~Kramer, B.~Mukherjee, and G.~Pesavento, ``{IPACT: A dynamic protocol for an
  Ethernet PON (EPON)},'' \emph{IEEE Communications Magazine}, vol.~40, no.~2,
  pp. 74--80, February 2002.

\bibitem{AYDA1103}
C.~Assi, Y.~Ye, S.~Dixit, and M.~Ali, ``Dynamic bandwidth allocation for
  {Quality-of-Service} over {Ethernet PONs},'' \emph{IEEE Journal on Selected
  Areas in Communications}, vol.~21, no.~9, pp. 1467--1477, November 2003.

\bibitem{MZC0303}
M.~Ma, Y.~Zhu, and T.~Cheng, ``{A bandwidth guaranteed polling MAC protocol for
  Ethernet passive optical networks},'' in \emph{Proceedings of IEEE INFOCOM},
  vol.~1, March 2003, pp. 22--31, {San Francisco, CA}.

\bibitem{NM0806}
H.~Naser and H.~Mouftah, ``A joint-{ONU} interval-based dynamic scheduling
  algorithm for ethernet passive optical networks,'' \emph{IEEE/ACM
  Transactions on Networking}, vol.~14, no.~4, pp. 889--899, August 2006.

\bibitem{ZhMo09}
J.~Zheng and H.~Mouftah, ``A survey of dynamic bandwidth allocation algorithms
  for {Ethernet Passive Optical Networks},'' \emph{Optical Switching and
  Networking}, vol.~6, no.~3, pp. 151--162, Jul. 2009.

\bibitem{Tak86}
H.~Takagi, \emph{Analysis of Polling Systems}.\hskip 1em plus 0.5em minus
  0.4em\relax MIT Press, 1986.

\bibitem{PaHR05}
C.~G. Park, D.~H. Han, and K.~W. Rim, ``Packet delay analysis of symmetric
  gated polling system for {DBA} scheme in an {EPON},'' \emph{Telecommunication
  Systems}, vol.~30, no. 1-3, pp. 13--34, Nov. 2005.

\bibitem{Holmb06}
T.~Holmberg, ``{Analysis of EPONs under the static priority scheduling scheme
  with fixed transmission times},'' in \emph{Proceedings of IEEE Conference on
  Next Generation Internet Design and Engineering (NGI)}, Apr. 2006, pp.
  192--199.

\bibitem{LaVC07}
B.~Lannoo, L.~Verslegers, D.~Colle, M.~Pickavet, M.~Gagnaire, and P.~Demeester,
  ``Analytical model for the {IPACT} dynamic bandwidth allocation algorithm in
  {EPONs},'' \emph{OSA Journal of Optical Networking}, vol.~6, no.~6, pp.
  677--688, Jun. 2007.

\bibitem{AuSHMR08}
F.~Aurzada, M.~Scheutzow, M.~Herzog, M.~Maier, and M.~Reisslein, ``Delay
  analysis of {Ethernet Passive Optical Networks} with gated service,''
  \emph{OSA Journal of Optical Networking}, vol.~7, no.~1, pp. 25--41, Jan.
  2008.

\bibitem{BaSY07}
X.~Bai, A.~Shami, and Y.~Ye, ``Delay analysis of {Ethernet} passive optical
  networks with quasi-leaved polling and gated service scheme,'' in
  \emph{Proc.\ of Second Int.\ Conference on Access Networks}, 2007, pp. 1--8.

\bibitem{BaS09}
S.~Bharati and P.~Saengudomlert, ``Packet delay analysis for limited service
  bandwidth allocation algorithm in {EPONs},'' in \emph{Proc.\ First Asian
  Himalayas International Conference on Internet (AH-ICI)}, 2009, pp. 1--5.

\bibitem{BhGB06}
S.~Bhatia, D.~Garbuzov, and R.~Bartos, ``{Analysis of the Gated IPACT Scheme
  for EPONs},'' in \emph{Proceedings of IEEE ICC}, Jun. 2006, pp. 2693--2698.

\bibitem{NgGB08}
M.~T. Ngo, A.~Gravey, and D.~Bhadauria, ``A mean value analysis approach for
  evaluating the performance of {EPON} with {Gated IPACT},'' in \emph{Proc.\ of
  Int.\ Conference on Optical Network Design and Modeling (ONDM)}, 2008, pp.
  1--6.

\bibitem{VaLo09}
J.~Vardakas and M.~Logothetis, ``Packet delay analysis for priority-based
  passive optical networks,'' in \emph{Proceedings of Int. Conf. on Emerging
  Network Intelligence}, Oct. 2009, pp. 103--107.

\bibitem{LuAn05icc}
Y.~Luo and N.~Ansari, ``{Dynamic upstream bandwidth allocation over Ethernet
  PONs},'' in \emph{Proceedings of IEEE ICC}, May 2005, pp. 1853--1857.

\bibitem{ZhMa08}
Y.~Zhu and M.~Ma, ``{IPACT} with grant estimation {(IPACT-GE)} scheme for
  {Ethernet} passive optical networks,'' \emph{IEEE/OSA Journal of Lightwave
  Technology}, vol.~26, no.~14, pp. 2055--2063, Jul. 2008.

\bibitem{VaVL09}
J.~Vardakas, V.~Vassilakis, and M.~Logothetis, ``Blocking analysis for priority
  classes in hybrid {WDM-OCDMA} passive optical networks,'' in \emph{Proc.\ of
  Fifth Advanced International Conference on Telecommunications (AICT)}, 2009,
  pp. 389--394.

\bibitem{Chang08}
W.-R. Chang, ``Research and design of system architectures and communication
  protocols for next-generation optical networks,'' Ph.D. dissertation,
  National Cheng Kung University, Tainan, Taiwan, 2008.

\bibitem{AuSRM10}
F.~Aurzada, M.~Scheutzow, M.~Reisslein, and M.~Maier, ``Towards a fundamental
  understanding of the stability and delay of offline {WDM EPONs},''
  \emph{IEEE/OSA Journal of Optical Communications and Networking}, vol.~2,
  no.~1, pp. 51--66, Jan. 2010.

\bibitem{CaFC00}
J.~Cai, A.~Fumagalli, and I.~Chlamtac, ``{The Multitoken Interarrival Time
  (MTIT) Access Protocol for Supporting Variable Size Packets Over WDM Ring
  Network},'' \emph{IEEE Journal on Selected Areas in Communications}, vol.~18,
  no.~10, pp. 2094--2104, Oct. 2000.

\bibitem{MBLM+96}
M.~A. Marsan, A.~Bianco, E.~Leonardi, M.~Meo, and F.~Neri, ``{On the Capacity
  of MAC Protocols for All--Optical WDM Multi--Rings with Tunable Transmitters
  and Fixed Receivers},'' in \emph{Proc., IEEE INFOCOM}, vol.~3, Mar. 1996, pp.
  1206--1216.

\bibitem{MLMN00}
M.~A. Marsan, E.~Leonardi, M.~Meo, and F.~Neri, ``{Modeling slotted WDM rings
  with discrete--time Markovian models},'' \emph{Computer Networks}, vol.~32,
  no.~5, pp. 599--615, May 2000.

\bibitem{RW96}
I.~Rubin and H.-T. Wu, ``Performance analysis and design of {CQBT} algorithm
  for a ring network with spatial reuse,'' \emph{IEEE/ACM Transactions on
  Networking}, vol.~4, no.~4, pp. 649--659, Aug. 1996.

\bibitem{ScRMS08}
M.~Scheutzow, M.~Reisslein, M.~Maier, and P.~Seeling, ``Multicast capacity of
  packet-switched ring {WDM} networks,'' \emph{IEEE Transactions on Information
  Theory}, vol.~54, no.~2, pp. 623--644, Feb. 2008.

\bibitem{DaGj04}
F.~Davik and S.~Gjessing, ``{The Stability of the Resilient Packet Ring
  Aggressive Fairness Algorithm},'' in \emph{Proc., IEEE LANMAN}, Apr. 2004,
  pp. 17--22.

\bibitem{GYBL04}
V.~Gambiroza, P.~Yuan, L.~Balzano, Y.~Liu, S.~Sheafor, and E.~Knightly,
  ``{Design, Analysis, and Implementation of DVSR: A Fair High-Performance
  Protocol for Packet Rings},'' \emph{IEEE/ACM Transactions on Networking},
  vol.~12, no.~1, pp. 85--102, Feb. 2004.

\bibitem{RH97}
I.~Rubin and H.-K.~H. Hua, ``{Synthesis and Throughput Behavior of WDM
  Meshed--Ring Networks Under Nonuniform Traffic Loading},'' \emph{IEEE/OSA
  Journal of Lightwave Technology}, vol.~15, no.~8, pp. 1513--1521, Aug. 1997.

\bibitem{Mehr90}
N.~Mehravari, ``Performance and protocol improvements for very high speed
  optical fiber local area networks using a passive star topology,''
  \emph{IEEE/OSA Journal of Lightwave Technology}, vol.~8, no.~4, pp. 520--530,
  Apr. 1990.

\bibitem{CHNS00}
N.~P. Caponio, A.~M. Hill, F.~Neri, and R.~Sabella, ``Single--layer optical
  platform based on {WDM/TDM} multiple access for large--scale 'switchless'
  networks,'' \emph{European Trans. on Telecomm.}, vol.~11, no.~1, pp. 73--82,
  Jan./Feb. 2000.

\bibitem{MaRe04}
M.~Maier and M.~Reisslein, ``{AWG}-based metro {WDM} networking,'' \emph{IEEE
  Communications Magazine}, vol.~42, no.~11, pp. S19--S26, Nov. 2004.

\bibitem{ScMRW03}
M.~Scheutzow, M.~Maier, M.~Reisslein, and A.~Wolisz, ``Wavelength reuse for
  efficient packet-switched transport in an {AWG}-based metro {WDM} network,''
  \emph{IEEE/OSA Journal of Lightwave Technology}, vol.~21, no.~6, pp.
  1435--1455, Jun. 2003.

\bibitem{YaMRC03}
H.-S. Yang, M.~Maier, M.~Reisslein, and W.~Carlyle, ``A genetic algorithm-based
  methodology for optimizing multiservice convergence in a metro {WDM}
  network,'' \emph{IEEE/OSA Journal of Lightwave Technology}, vol.~21, no.~5,
  pp. 1114--1133, May 2003.

\bibitem{ChKS06}
C.-H. Chang, P.~Kourtessis, and J.~M. Senior, ``{GPON} service level agreement
  based dynamic bandwidth assignment protocol,'' \emph{IEE Electronics
  Letters}, vol.~42, no.~20, pp. 1173--1174, Sep. 2006.

\bibitem{JiHS06}
J.~Jiang, M.~R. Handley, and J.~M. Senior, ``Dynamic bandwidth assignment {MAC}
  protocol for differentiated services over {GPON},'' \emph{IEE Electronics
  Letters}, vol.~42, no.~11, pp. 653--655, May 2006.

\bibitem{JiSe09}
J.~Jiang and J.~Senior, ``A new efficient dynamic {MAC} protocol for the
  delivery of multiple services over {GPON},'' \emph{Photonic Network
  Communications}, vol.~18, no.~2, pp. 227--236, Oct. 2009.

\bibitem{HaSM06}
M.~Hajduczenia, H.~J.~A. da~Silva, and P.~P. Monteiro, ``{EPON} versus {APON}
  and {GPON}: a detailed performance comparison,'' \emph{OSA Journal of Optical
  Networking}, vol.~5, no.~4, pp. 298--319, Apr. 2006.

\bibitem{SCAW09}
B.~Skubic, J.~Chen, J.~Ahmed, L.~Wosinska, and B.~Mukherjee, ``A comparison of
  dynamic bandwidth allocation for {EPON}, {GPON}, and next-generation {TDM
  PON},'' \emph{IEEE Communications Magazine}, vol.~47, no.~3, pp. S40--S48,
  Mar. 2009.

\bibitem{McMR06}
M.~McGarry, M.~Maier, and M.~Reisslein, ``{WDM} {Ethernet} passive optical
  networks,'' \emph{IEEE Comm.\ Mag.}, vol.~44, no.~2, pp. S18--S25, Feb. 2006.

\bibitem{DaGR09}
R.~Davey, D.~Grossman, M.~Rasztovits-Wiech, D.~Payne, D.~Nesset, A.~Kelly,
  A.~Rafel, S.~Appathurai, and S.-H. Yang, ``Long-reach passive optical
  networks,'' \emph{IEEE/OSA Journal of Lightwave Technology}, vol.~27, no.~3,
  pp. 273--291, Feb. 2009.

\bibitem{LaPC08}
J.~Lazaro, J.~Prat, P.~Chanclou, G.~T. Beleffi, A.~Teixeira, I.~Tomkos,
  R.~Soila, and V.~Koratzinos, ``Scalable extended reach {PON},'' in
  \emph{Proc. OFC}, Feb. 2008.

\bibitem{SCLGLBP0909}
F.~Saliou, P.~Chanclou, F.~Laurent, N.~Genay, J.~Lazaro, F.~Bonada, and
  J.~Prat, ``{Reach Extension Strategies for Passive Optical Networks},''
  \emph{IEEE/OSA Journal of Optical Communications and Networking}, vol.~1,
  no.~4, pp. C51--C60, Sep 2009.

\bibitem{SM0209}
D.~Shea and J.~Mitchell, ``Architecture to integrate multiple {PONs} with long
  reach {DWDM} backhaul,'' \emph{IEEE Journal on Selected Areas in
  Communications}, vol.~27, no.~2, pp. 126--133, Feb 2009.

\bibitem{TaTo06}
G.~Talli and P.~D. Townsend, ``Hybrid {DWDM-TDM} long-reach {PON} for
  next-generation optical access,'' \emph{IEEE/OSA Journal of Lightwave
  Technology}, vol.~24, no.~7, pp. 2827--2834, Jul. 2006.

\bibitem{MADM09}
L.~Meng, C.~Assi, M.~Maier, and A.~Dhaini, ``Resource management in
  {STARGATE}-based {Ethernet} passive optical networks {(SG-EPONs)},'' in
  \emph{Proc. IEEE ICC}, Dresden, Germany, Jun. 2009.

\bibitem{MHSR05}
M.~Maier, M.~Herzog, M.~Scheutzow, and M.~Reisslein, ``{PROTECTORATION}: A fast
  and efficient multiple-failure recovery technique for resilient packet ring
  {(RPR)} using dark fiber,'' \emph{IEEE/OSA Journal of Lightwave Technology},
  vol.~23, no.~10, pp. 2816--2838, Oct. 2005.

\bibitem{Klein75}
L.~Kleinrock, \emph{Queueing Systems: Volume I: Theory}.\hskip 1em plus 0.5em
  minus 0.4em\relax Wiley, 1975.

\bibitem{BuS83}
W.~Bux and M.~Schlatter, ``An approximate method for the performance analysis
  of buffer insertion rings,'' \emph{IEEE Transactions on Communications}, vol.
  COM-31, no.~1, pp. 50--55, Jan. 1983.

\end{thebibliography}


\end{document}